\documentclass[useAMS, usenatbib]{mnras}

\usepackage[T1]{fontenc}

\DeclareRobustCommand{\VAN}[3]{#2}
\let\VANthebibliography\thebibliography
\def\thebibliography{\DeclareRobustCommand{\VAN}[3]{##3}\VANthebibliography}

\usepackage{graphicx}	
\usepackage{amsmath}	
\usepackage{amssymb}	
\usepackage[export]{adjustbox}
\usepackage{subcaption}
\usepackage{epsfig}
\usepackage{newtxtext,newtxmath}
\usepackage{booktabs}

\newcommand{\ms}{\ifmmode{\,M_{\odot}}\else{\,$M_{\odot}$}\fi}
\newcommand{\kms}{\ifmmode{\mathrm{\,km\,s^{-1}}}\else{\,$\mathrm{km\,s^{-1}}$}\fi}
\defcitealias{2023ApJ...956..132L}{L23}

\title[Optical counterpart of a redback MSP]{Discovery of the variable optical counterpart of the redback pulsar PSR~J2055$+$1545}

\author[Turchetta et al.]{Marco Turchetta,$^{1}$\thanks{E-mail: marcoturchetta@ntnu.no} Bidisha Sen$^{1}$, Jordan A. Simpson$^{1}$, Manuel Linares$^{1,2}$,\newauthor Rene P. Breton$^{3}$, Jorge Casares$^{4,5}$, Mark R. Kennedy$^{3,6}$ and Tariq Shahbaz$^{4,5}$
\\
$^{1}$ Department of Physics, Norwegian University of Science and Technology, NO-7491 Trondheim, Norway\\
$^{2}$ Departament de F{\'i}sica, EEBE, Universitat Polit{\`e}cnica de Catalunya, Av. Eduard Maristany 16, E-08019 Barcelona, Spain\\
$^{3}$ Jodrell Bank Centre for Astrophysics, Department of Physics and Astronomy, The University of Manchester, Manchester M13 9PL, UK\\
$^{4}$ Instituto de Astrofísica de Canarias, E-38205 La Laguna, Tenerife, Spain\\
$^{5}$ Departamento de Astrofísica, Universidad de La Laguna, E-38206 La Laguna, Tenerife, Spain\\
$^{6}$ School of Physics, University College Cork, Cork, Ireland.
}

\date{Accepted XXX. Received YYY; in original form ZZZ}

\pubyear{2024}

\begin{document}
\label{firstpage}
\pagerange{\pageref{firstpage}--\pageref{lastpage}}
\maketitle

\begin{abstract}
We present the discovery of the variable optical counterpart to PSR~J2055$+$1545, a redback millisecond pulsar, and the first radial velocity curve of its companion star. The multi-band optical light curves of this system show a $0.4$--$0.6 \ \mathrm{mag}$ amplitude modulation with a single peak per orbit and variable colours, suggesting that the companion is mildly irradiated by the pulsar wind.
We find that the flux maximum is asymmetric and occurs at orbital phase $\simeq0.4$, anticipating the superior conjunction of the companion (where the optical emission of irradiated redback companions is typically brightest). We ascribe this asymmetry, well fit with a hot spot in our light curve modelling, to irradiation from the intrabinary shock between pulsar and companion winds.
%
The optical spectra obtained with the \textit{Gran Telescopio Canarias} reveal a G-dwarf companion star with temperatures of $5749 \pm 34 \ \mathrm{K}$ and $6106 \pm 35 \ \mathrm{K}$ at its inferior and superior orbital conjunctions, respectively, and a radial velocity semi-amplitude of $385 \pm 3 \kms{}$.
%
Our best-fit model yields a neutron star mass of $1.7^{+0.4}_{-0.1} \ \mathrm{M}_{\sun}$ and a companion mass of $0.29^{+0.07}_{-0.01} \ \mathrm{M}_{\sun}$. Based on the close similarity between the optical light curve of PSR~J2055$+$1545 and those observed from PSR~J1023$+$0038 and PSR~J1227$-$4853 during their rotation-powered states, we suggest this system may develop an accretion disc in the future and manifest as a transitional millisecond pulsar.
\end{abstract}

\begin{keywords}
 techniques: photometric -- binaries: close -- stars: neutron -- pulsars: general -- pulsars: individual: PSR J2055+1545 
\end{keywords}



\section{Introduction}
\label{sec:intro}

Millisecond pulsars (MSPs) are rapidly rotating neutron stars that originate from low-mass X-ray binary (LMXB) progenitor systems. During this LMXB phase, the companion star transfers matter and angular momentum to the neutron star, spinning it up to $\sim\mathrm{ms}$ periods \citep{1982CSci...51.1096R}. Once the outward pressure exerted by the neutron star's magnetic field exceeds the inward ram pressure of the accreting material, the accretion disc is expelled, allowing the neutron star to become active as a rotation-powered radio MSP \citep{2013A&A...558A..39T}.

\textit{Spiders} are a sub-class of MSPs found in compact binary systems ($P_{\mathrm{orb}}\lesssim1 \ \mathrm{d}$), where the pulsar's high-energy particle wind can irradiate and progressively consume their companion stars, potentially fully devouring them in few cases \citep{1988Natur.333..237F,1988Natur.334..227V}. This extreme interaction has earned these systems the nicknames \textit{black widows} (BWs) and \textit{redbacks} (RBs), characterized by companion masses $M_{2}\sim 0.01 \ \mathrm{M}_{\sun}$ and $M_{2}\simeq 0.3$--$0.7 \ \mathrm{M}_{\sun}$, respectively. Notably, three transitional MSPs within the RB category have been observed switching between the accretion-powered disc state and the radio pulsar state over timescales of a few weeks to months \citep{2009Sci...324.1411A,2013Natur.501..517P,2014MNRAS.441.1825B}, positioning RBs as the missing link between the LMXB and MSP evolutionary phases. Furthermore, spider binaries in general are considered prime systems for hosting super-massive neutron stars, having undergone prolonged Gyr-long accretion episodes before activating as radio MSPs \citep{2020mbhe.confE..23L}.

Spider MSPs often show large eclipses of their radio pulsations, caused by absorption from material expelled by the companion star \citep{2020MNRAS.494.2948P}, making their detection in blind radio surveys particularly challenging. Over the past 16 years, the \textit{Fermi Large Area Telescope} \citep[\textit{Fermi}-LAT;][]{2020ApJS..247...33A} has played a crucial role in expanding the known spider population, with 65 confirmed systems to date \citep{nedreaas2024spidercat}. Since its launch, \textit{Fermi} has enabled the discovery of numerous spider MSPs, not only as $\gamma$-ray emitters \citep{2023ApJ...958..191S}, but also through targeted radio searches of previously unassociated $\gamma$-ray sources \citep{2024MNRAS.530.4676T}.

Spider systems can also be identified through the variable optical emission of their companion stars \citep[e.g.][]{2015ApJ...814...88S, 2017MNRAS.465.4602L, 2023ApJ...943..103A}. The shape of their light curves is primarily determined by the degree of heating from the pulsar wind, which in turn depends on the system's orbital period, the companion star's intrinsic `base' temperature and the pulsar's spin-down luminosity \citep{2023MNRAS.525.2565T}. Indeed, all currently known BWs exhibit irradiation-dominated optical light curves, with peak-to-peak amplitudes $\gtrsim1 \ \mathrm{mag}$ and a single flux maximum per orbit, owing to their relatively cold companion stars ($\simeq1000$--$3000 \ \mathrm{K}$). On the other hand RB companions are hotter ($\simeq4000$--$6000 \ \mathrm{K}$), thus about half of them show little to no irradiation effects, resulting in double-peaked light curves with amplitudes $\simeq0.3 \ \mathrm{mag}$ driven by the tidal distortion of the companion. Combining optical light curve modelling with independent spectroscopic measurements of the companion's radial velocity is crucial for obtaining precise and robust constraints on the fundamental parameters of spider systems, particularly their neutron star masses \citep[e.g.][]{2018ApJ...859...54L,2021ApJ...908L..46R,2024MNRAS.528.4337D}.

PSR~J2055$+$1545 (hereafter referred to as J2055) is a RB system discovered as a radio MSP by \cite{2023ApJ...956..132L}, with a spin period of $2.16 \ \mathrm{ms}$ and an orbital period of $4.8 \ \mathrm{hr}$. It has been classified as a RB MSP based on the observed radio eclipse, spanning $36\%$ of its orbit, and an estimated median companion mass of $0.29 \ \mathrm{M}_{\sun}$ \citep{2023ApJ...956..132L}. Additionally, J2055 is associated with the pulsar-like \textit{Fermi} source 4FGL~J2055.8$+$1545, although no $\gamma$-ray pulsations have been detected. This is likely due to the system’s substantial orbital variability, which makes it difficult to maintain phase coherence outside the time range covered by the radio ephemeris. Optical counterparts to J2055 have been identified in \textit{Gaia} Data Release 3 \citep[DR3;][]{refId0}, \textit{Pan-STARRS}~1 \citep[PS1;][]{2016arXiv161205560C}, and \textit{Sloan Digital Sky Survey} Data Release 17 \citep[SDSS-DR17;][]{Abdurro'uf_2022}, no optical photometric or spectroscopic observations have yet been performed. Consequently, key parameters such as the neutron star mass, orbital inclination, and companion temperature remain either unknown or poorly constrained.

In this paper, we present the discovery of the variable optical counterpart of J2055 and the first radial velocity curve for this system. Section \ref{sec:obsanddata} details the two optical campaigns conducted in 2023 and 2024, along with the photometric and spectral analysis techniques employed. In Section \ref{sec:results}, we report the optical light curves of J2055, which exhibit a single asymmetric flux maximum per orbit, indicative of an irradiated companion, as well as the system's radial velocity curve. Section \ref{sec:modelling} focuses on the combined modelling of the 2023 and 2024 light curves using the {\sc Icarus} software \citep{2012ApJ...748..115B}, where we use the semi-amplitude of the companion's radial velocity, derived from spectroscopy, as the central value for its prior distribution. In Section \ref{sec:discussion}, we interpret and discuss our findings, comparing the key optical features and the intermediate irradiated regime observed in J2055 with those of other RB systems.

\section{Observations and data analysis}
\label{sec:obsanddata}
\subsection{Photometry}
\label{subsec:photometry}
\begin{table*}
    \centering
    \caption{Log of the optical photometric campaign for J2055, with data acquired using the LCO global network of $1$-$\mathrm{m}$ telescopes and the $2.56$-$\mathrm{m}$ NOT telescope.}\label{tab:photlog}
    \setlength{\tabcolsep}{10.5pt}
    \begin{tabular}{ccccccc}
    \\[-1.8ex]\hline
    \hline \\[-1.8ex]
        Telescope & Instrument & Date & Time & Images in \textit{g'} & Images in \textit{r'} & Images in \textit{i'} \\
        (diameter) & (configuration) & (evening) & (UT) & ($\mathrm{nr}\times\mathrm{exp.}\ \mathrm{time}$) & ($\mathrm{nr}\times\mathrm{exp.}\ \mathrm{time}$) & ($\mathrm{nr}\times\mathrm{exp.}\ \mathrm{time}$) \\[1.0ex] 
        \hline\\[-1.8ex]
         LCO Siding Spring-1m & Sinistro-1x1 & 2023-07-25 & 13:29$-$16:18 & $10\times300 \ \mathrm{s}$ & $10\times300 \ \mathrm{s}$ & $10\times300 \ \mathrm{s}$ \\
         LCO McDonald-1m & Sinistro-1x1 & 2023-07-25 & 08:03$-$10:51 & $10\times300 \ \mathrm{s}$ & $10\times300 \ \mathrm{s}$ & $10\times300 \ \mathrm{s}$ \\
         LCO Siding Spring-1m & Sinistro-1x1 & 2023-07-28 & 12:31$-$15:20 & $10\times300 \ \mathrm{s}$ & $10\times300 \ \mathrm{s}$ & $10\times300 \ \mathrm{s}$ \\
         LCO McDonald-1m & Sinistro-1x1 & 2023-07-28 & 06:04$-$08:53 & $9\times300 \ \mathrm{s}$ & $10\times300 \ \mathrm{s}$ & $10\times300 \ \mathrm{s}$ \\
         LCO McDonald-1m & Sinistro-1x1 & 2023-07-29 & 06:00$-$08:49 & $10\times300 \ \mathrm{s}$ & $10\times300 \ \mathrm{s}$ & $10\times300 \ \mathrm{s}$ \\
         LCO McDonald-1m & Sinistro-1x1 & 2023-09-10 & 04:17$-$07:05 & $10\times300 \ \mathrm{s}$ & $10\times300 \ \mathrm{s}$ & $10\times300 \ \mathrm{s}$ \\
         LCO Sutherland-1m & Sinistro-1x1 & 2023-09-13 & 18:55$-$19:11 & $2\times300 \ \mathrm{s}$ & $2\times300 \ \mathrm{s}$ & $1\times300 \ \mathrm{s}$ \\
         LCO Sutherland-1m & Sinistro-1x1 & 2023-09-14 & 18:55$-$21:10 & $4\times300 \ \mathrm{s}$ & $6\times300 \ \mathrm{s}$ & $5\times300 \ \mathrm{s}$ \\
         LCO McDonald-1m & Sinistro-1x1 & 2023-09-14 & 04:55$-$07:44 & $10\times300 \ \mathrm{s}$ & $10\times300 \ \mathrm{s}$ & $10\times300 \ \mathrm{s}$ \\
         LCO McDonald-1m & Sinistro-1x1 & 2023-09-15 & 03:38$-$05:53 & $8\times300 \ \mathrm{s}$ & $8\times300 \ \mathrm{s}$ & $8\times300 \ \mathrm{s}$ \\
         LCO McDonald-1m & Sinistro-1x1 & 2023-09-17 & 04:33$-$07:21 & $10\times300 \ \mathrm{s}$ & $10\times300 \ \mathrm{s}$ & $10\times300 \ \mathrm{s}$ \\
         LCO McDonald-1m & Sinistro-1x1 & 2023-09-18 & 04:45$-$06:42 & $7\times300 \ \mathrm{s}$ & $7\times300 \ \mathrm{s}$ & $6\times300 \ \mathrm{s}$ \\
         LCO Teide-1m & Sinistro-1x1 & 2023-10-10 & 22:00$-$22:33 & $3\times300 \ \mathrm{s}$ & $3\times300 \ \mathrm{s}$ & $2\times300 \ \mathrm{s}$ \\
         LCO McDonald-1m & Sinistro-1x1 & 2023-10-10 & 02:35$-$05:23 & $10\times300 \ \mathrm{s}$ & $10\times300 \ \mathrm{s}$ & $10\times300 \ \mathrm{s}$ \\
         LCO McDonald-1m & Sinistro-1x1 & 2023-10-11 & 03:13$-$06:01 & $7\times300 \ \mathrm{s}$ & $8\times300 \ \mathrm{s}$ & $8\times300 \ \mathrm{s}$ \\
         LCO Teide-1m & Sinistro-1x1 & 2023-10-12 & 20:27$-$22:42 & $8\times300 \ \mathrm{s}$ & $8\times300 \ \mathrm{s}$ & $8\times300 \ \mathrm{s}$ \\
         LCO Teide-1m & Sinistro-1x1 & 2023-10-13 & 21:49$-$00:03 & $8\times300 \ \mathrm{s}$ & $8\times300 \ \mathrm{s}$ & $8\times300 \ \mathrm{s}$ \\
         LCO McDonald-1m & Sinistro-1x1 & 2023-10-13 & 02:53$-$05:42 & $10\times300 \ \mathrm{s}$ & $10\times300 \ \mathrm{s}$ & $10\times300 \ \mathrm{s}$ \\
         LCO McDonald-1m & Sinistro-1x1 & 2023-10-15 & 02:42$-$04:57 & $8\times300 \ \mathrm{s}$ & $8\times300 \ \mathrm{s}$ & $8\times300 \ \mathrm{s}$ \\
         NOT-2.56m & ALFOSC-2x2 & 2023-10-18 & 21:51$-$00:56 & $9\times300 \ \mathrm{s}$ & $18\times300 \ \mathrm{s}$ & $9\times300 \ \mathrm{s}$ \\
         NOT-2.56m & ALFOSC-2x2 & 2023-11-04 & 21:29$-$23:31 & $6\times300 \ \mathrm{s}$ & $12\times300 \ \mathrm{s}$ & $6\times300 \ \mathrm{s}$ \\
         NOT-2.56m & ALFOSC-2x2 & 2024-08-01 & 22:58$-$03:37 & $18\times300 \ \mathrm{s}$ & $18\times300 \ \mathrm{s}$ & $18\times300 \ \mathrm{s}$ \\
        \hline
    \end{tabular}
    \label{tab:photlog}
\end{table*}

We carried out an optical photometric campaign of J2055 spanning one year, from July 2023 to August 2024. The 2023 observations were obtained using the Sinistro and ALFOSC cameras mounted on the $1$-$\mathrm{m}$ LCO and $2.56$-$\mathrm{m}$ NOT telescopes, respectively, while the 2024 data were exclusively acquired with the NOT/ALFOSC setup. The observations consisted of 5-min long exposures, alternating between the SDSS \textit{g'}, \textit{r'} and \textit{i'} filters. Instrumental setups and additional details of these observations are provided in Table \ref{tab:photlog}.

We processed the LCO and NOT data using the {\sc banzai} data-reduction pipeline\footnote{https://lco.global/documentation/data/BANZAIpipeline/} and the standard tools of {\sc IRAF} \citep[v. 2.16,][]{IRAF}, respectively. The photometric techniques applied to the data were the same for LCO and NOT images. The analyses were performed separately on the 2023 LCO, 2023 NOT, and 2024 NOT datasets.

\begin{figure}
    \centering
    \includegraphics[width=\columnwidth]{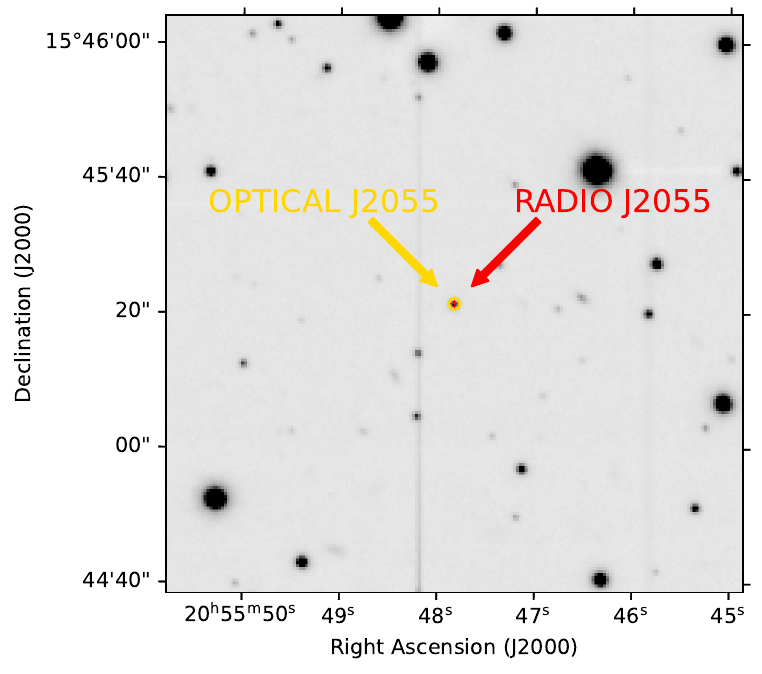}
    \caption{Median image in the \textit{r'}-band obtained from the NOT/ALFOSC dataset acquired in 2024. The radio and optical locations of J2055 are highlighted in red and yellow, respectively.}
    \label{fig:J2055FoV}
\end{figure}

We combined the reduced images into a single median image for each of the three optical bands, \textit{g'}, \textit{r'} and \textit{i'}, to enhance the source detection sensitivity. We identified all the sources having a signal-to-noise ratio $\geq3$ and $\geq2$ in these median frames for LCO and NOT, respectively, using the {\sc SEP}\footnote{\url{https://github.com/kbarbary/sep}} package \citep{2016zndo....159035B}, which is based on the {\sc SExtractor} software \citep{1996A&AS..117..393B}. Figure \ref{fig:J2055FoV} shows the median image in the \textit{r'}-band obtained from 2024 NOT data, centred around the coinciding radio and optical locations of J2055, marked in red and yellow, respectively.

We performed systematic aperture photometry on all the detected sources using {\sc SEP}, with aperture radii of $1.2$ and $1.5 \times$ the average full-width at half maximum (FWHM) for the LCO and NOT data, respectively. To improve the accuracy of our photometry, we utilized the {\sc ASTROSOURCE} package\footnote{\url{https://github.com/zemogle/astrosource}} \citep{2021JOSS....6.2641F} to select the 8--10 most stable and brightest stars for each optical filter (\textit{g'}, \textit{r'}, and \textit{i'}) and dataset (2023 LCO, 2023 NOT, and 2024 NOT). We used these stars, with variability rms amplitudes of $\simeq 0.004$ – $0.01 \ \mathrm{mag}$ and magnitudes between 16 and 18.5, as comparison stars to obtain the highest-precision photometry for our target, J2055.

\subsection{Spectroscopy}
\label{subsec:spectroscopy}

We observed J2055 with the 10.4-m Gran Telescopio Canarias
(GTC) on the night of August 1st, 2024 (from MJD 60523.963626 to
60524.152712).
We obtained 18 OSIRIS+ long-slit spectra with an exposure time of
935~s each, at airmass 1.03-1.23 and under good seeing ($\simeq$0.6'')
and weather (clear sky) conditions.
We used the R2500V grism with a central wavelength of 5185{~\AA} and a
slit width of 1'', covering the 4500--6000{~\AA} spectral range.
Together, our spectra cover 4.5 uninterrupted hours, corresponding to $94\%$ of the
pulsar's orbit.

We applied standard calibration procedures (bias and spectroscopic
flat corrections) within \textsc{iraf}.
We optimally extracted sky-subtracted spectra
using \textsc{starlink/pamela} to account for significant
tilt \citep{Marsh89}.
We calibrated the resulting spectra in wavelength within
\textsc{molly}, using 18 identified lines from a collated set of
three arc spectra (Ne, HgAr and Xe lamps) taken on the same evening.
The wavelength scale was well-fit with a 4-term polynomial (r.m.s.
0.03\,\AA{}), with a central dispersion of about 0.80\,\AA{}\,px$^{-1}$.
From the 5577\,\AA{} (sky) and 5461\,\AA{} (arc) lines, we estimate a
spectral resolution of 3.1 and 3.0\,\AA{}, respectively (i.e., a
dimensionless resolution of $R \simeq 1800$).
\section{Results}
\label{sec:results}
\subsection{Optical light curves of the redback PSR~J2055$+$1545}
\label{sec:optlcJ2055}

\begin{figure*}
    \centering
    \includegraphics[width=0.49\textwidth]{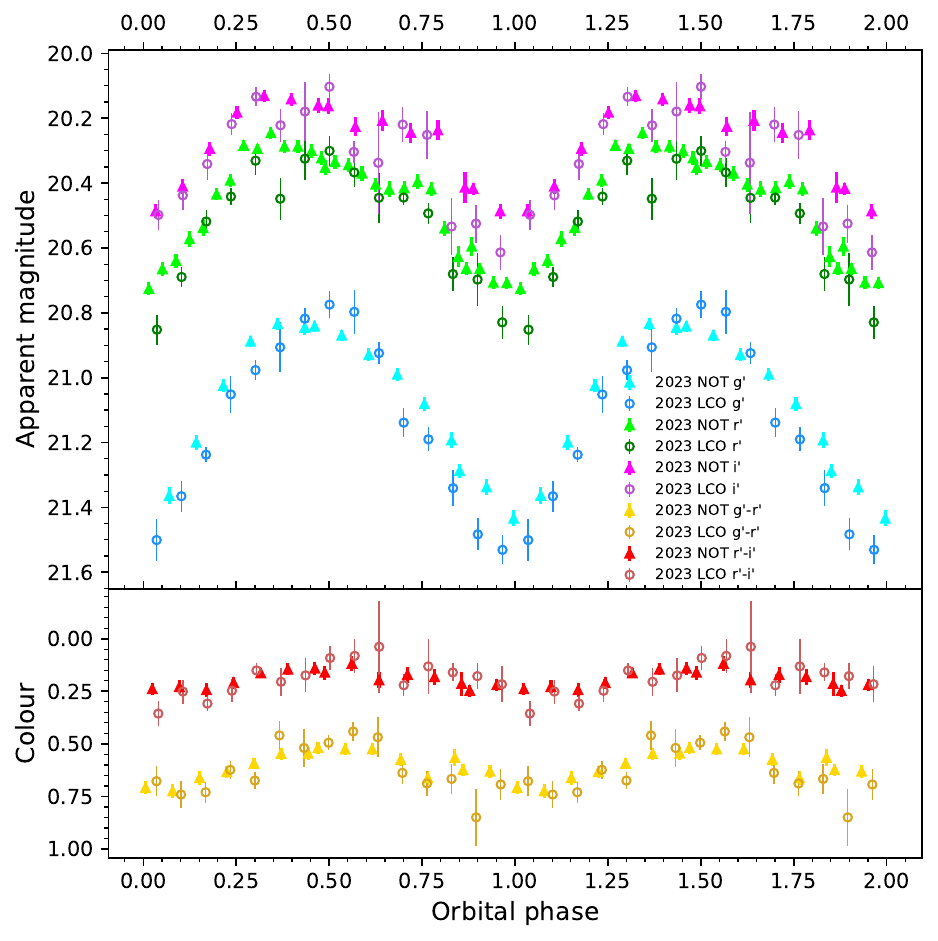}
    \includegraphics[width=0.49\textwidth]{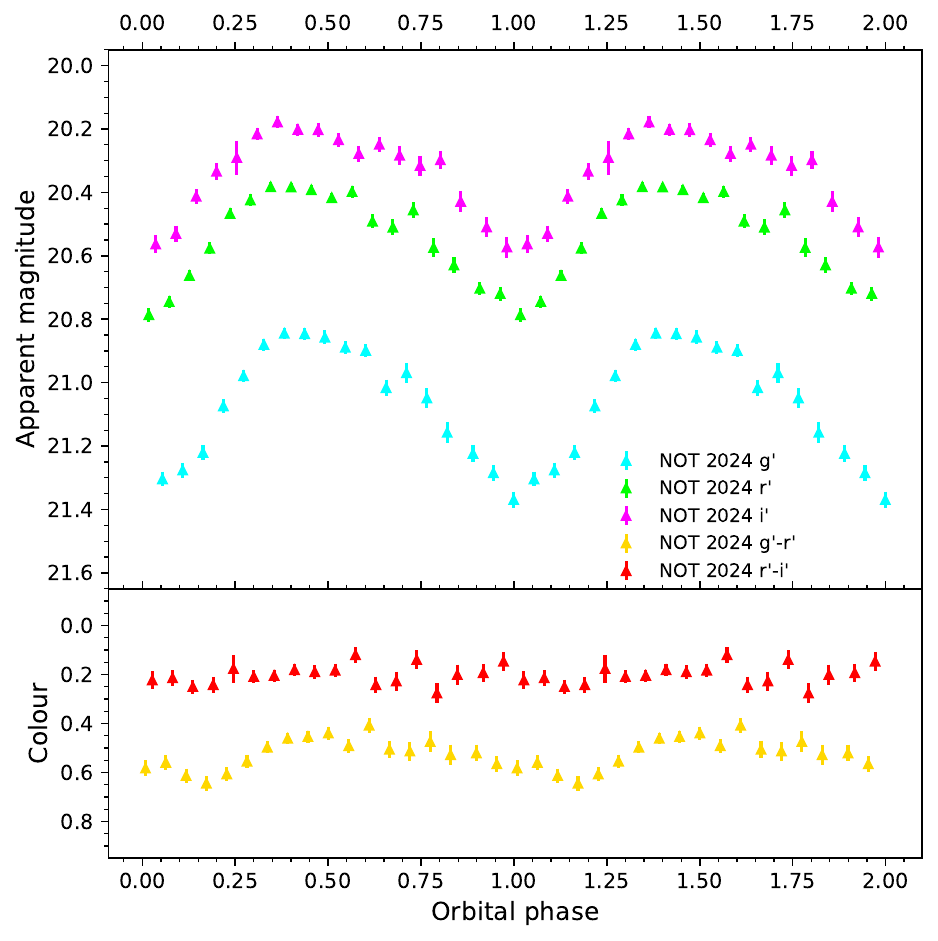}
    \caption{\textit{Top panels}: The optical light curves of J2055 in \textit{g'}, \textit{r'} and \textit{i'} bands, folded at the orbital period $P_\mathrm{orb} \ \mathrm{d}$ and reference epoch $T_\mathrm{0}=60523.9866 \ \mathrm{MJD}$, for data obtained in 2023 (top-left panel) and 2024 (top-right panel). \textit{Bottom panels}: The observed colours (\textit{g'} - \textit{r'}) and (\textit{r'} - \textit{i'}) for the same years, shown in the bottom-left and bottom-right panels, respectively. Data from LCO/Sinistro and NOT/ALFOSC are plotted with empty circles and filled triangles, respectively. Each panel displays two orbital cycles for clarity. The LCO/Sinistro light and colour curves from 2023 have been rebinned in phase for easier comparison with the NOT/ALFOSC data.}
    \label{fig:lcgri}
\end{figure*}

The mid-exposure UTC times of the frames were converted into the TDB coordinate system. To verify whether the orbital period of J2055 had significantly deviated from the known value of $P_\mathrm{orb}=0.200725452(1) \ \mathrm{d}$ \citep[][hereafter \citetalias{2023ApJ...956..132L}]{2023ApJ...956..132L}, we conducted a phase-dispersion minimization periodicity search \citep{1978ApJ...224..953S} on the LCO data\footnote{The timing solution of J2055 reported by \citetalias{2023ApJ...956..132L} shows a high degree of orbital variability, with eight orbital frequency derivative terms.}. The best photometric period obtained was $0.200725(5) \ \mathrm{d}$, which is fully consistent with the value found by \citetalias{2023ApJ...956..132L}. Therefore, we adopted their orbital period to phase-fold optical light curves and colours of J2055 both for 2023 NOT/LCO and 2024 NOT datasets (shown in the left and right panels of Figure \ref{fig:lcgri}, respectively). We set phase $0$ to correspond to the time of the companion's inferior conjunction, $T_\mathrm{0}=60523.9866(3) \ \mathrm{MJD}$, as determined from the optical spectra of J2055 (see Section \ref{sec:optrvcJ2055} for details). Given the larger size of the LCO dataset compared to NOT (see Table \ref{tab:photlog}), we phase-binned the LCO light and colour curves into 15 bins to facilitate comparison with the NOT observations. To do so, we computed the average magnitude over all the points in the corresponding phase bin, with their standard deviation as uncertainty.

As shown in Figure \ref{fig:lcgri}, the optical light curves of J2055 exhibit the same periodic modulation in both 2023 and 2024, with a single flux maximum per orbit and peak-to-peak amplitudes of $0.6$, $0.5$ and $0.4 \ \mathrm{mag}$ in the \textit{g'}, \textit{r'} and \textit{i'} bands, respectively. We also observe variable (\textit{g'} - \textit{r'}) and (\textit{r'} - \textit{i'}) colours along the orbit in both epochs, with maxima occurring around the companion's superior conjunction (phase $0.5$). After correcting for extinction the NOT colours observed at phase $0$ and $0.5$ using a colour excess of $E(g-r)=0.07\pm0.01$ \citep{2019ApJ...887...93G} and matching them to the low-mass spectral templates of \cite{allard11}, we estimated the companion temperatures at inferior and superior conjunctions, respectively\footnote{In previous work, the temperatures measured at the companion's inferior and superior conjunctions have often been referred to as `night-side' and `day-side' temperatures: $T_{\mathrm{night}}$ and $T_{\mathrm{day}}$, respectively \citep[e.g., see][]{2013ApJ...769..108B,2018ApJ...859...54L}. However, `night' and `day' should actually correspond to the temperatures of the companion's hemispheres opposite to and facing the pulsar, respectively. This differs from our observed $T_{\mathrm{inf}}$ and $T_{\mathrm{sup}}$, unless the orbital inclination is exactly 90$^\circ$ (i.e., unless the orbit is seen exactly edge-on).}.
For such temperature estimates, we used only data obtained with the $2.56$-$\mathrm{m}$ NOT telescope, as they have smaller magnitude errors compared to those from the $1$-$\mathrm{m}$ LCO.
From the 2023 data, we derived intrinsic colours of $(g'-r') = 0.66 \pm 0.05 \ \mathrm{mag}$ at phase $0$ and $(g'-r') = 0.45 \pm 0.05 \ \mathrm{mag}$ at phase $0.5$, corresponding to temperatures of $T_{\mathrm{inf}} = 5300 \pm 200 \ \mathrm{K}$ and $T_{\mathrm{sup}} = 5900 \pm 200 \ \mathrm{K}$, respectively. In the 2024 dataset, we found intrinsic colours of $(g'-r') = 0.50 \pm 0.05$ and $(g'-r') = 0.37 \pm 0.05$ at inferior and superior conjunctions, respectively, which correspond to temperatures of $T_{\mathrm{inf}} = 5700 \pm 175 \ \mathrm{K}$ and $T_{\mathrm{sup}} = 6175 \pm 200 \ \mathrm{K}$. These values are consistent with those derived from the 2023 data.

The features described above strongly suggest that the companion star of J2055 is mildly heated by the pulsar wind (see Section \ref{sec:intro}). The consistency between the 2023 and 2024 observations indicates that this irradiated state persisted from July 2023 to August 2024, without any state change in the system. Interestingly, the light curve shows a clearly asymmetric flux maximum at phase $\simeq 0.4$, anticipating the companion's superior conjunction, where the optical flux in irradiated spiders typically peaks. Additionally, a local maximum is observed at the descending node of the companion (phase $0.75$), which appears sharper in the \textit{r'} and \textit{i'} bands compared to the \textit{g'}-band. In Section \ref{sec:modelling}, we explain this complex behaviour by including a hot spot component in our optical light curve modelling. Subsequently, we discuss the implications of our results in the context of other observed RB systems (see Section \ref{sec:discussion}).

\subsection{Spectroscopic results from the redback PSR~J2055$+$1545}
\label{sec:optrvcJ2055}

\begin{figure}
    \centering
    \includegraphics{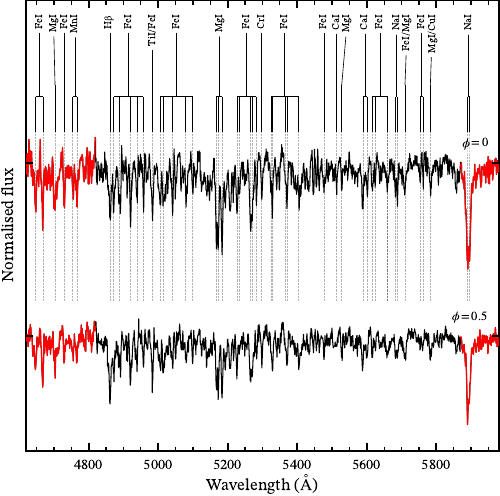}
    \caption{Average spectra from J2055 about companion inferior conjunction (top spectrum, $\phi=0$) and companion superior conjunction (bottom spectrum, $\phi=0.5$). Identifications for the most prominent lines are indicated above the spectra, based on the Solar atlas of \citet{solaratlas}. The wavelength range used for optimal subtraction is shown in black, while excluded ranges are highlighted in red. The `metals' radial velocity curve (Figure~\ref{fig:rvs}, green) was obtained using the same wavelength range, while also excluding the H$\beta$ line and the \ion{Mg}{i} triplet.}
    \label{fig:optsub}
\end{figure}

The optical spectra of J2055's companion are consistent with a relatively typical G-dwarf star, which is lightly irradiated by the pulsar wind. A myriad of metallic line features are visible throughout the spectra (see Figure~\ref{fig:optsub}), including the \ion{Mg}{i} triplet at 5167--5183\,\AA{} and various strong \ion{Fe}{i} and \ion{Ca}{i} lines from 4900 to 5400\,\AA{}. These metallic features vary in strength throughout the orbit, appearing strongest around phase $0$, when the cold face of the companion is visible. The H$\beta$ line is also a prominent feature, and shows a much more subtle, and reversed, trend -- peaking in strength at phase 0.5, when we see the hot face.

We measured the temperature of the companion throughout its orbit using optimal subtraction \citep{marsh94} between the normalised spectra of J2055 and a set of template spectra, as implemented by the \texttt{astra} package\footnote{\url{http://github.com/jsimpson-astro/astra}}. This optimises a scaling factor $f_\mathrm{star}$ to minimise residuals between the target and template spectra, accounting for contributions from non-stellar flux (which we define as $f_\mathrm{veil} = 1 - f_\mathrm{star}$). The spectra of J2055 were first shifted to zero velocity before being averaged in 4 phase bins of a quarter phase in width, centred around $\phi = 0$, 0.25, 0.5, and 0.75. We used a set of templates created from the BT-Settl library of synthetic spectra \citep{allard11} covering the temperature range $2600 \le T_\mathrm{eff} \le 10000$\,K, broadened to match the instrumental resolution. Thus, the results of optimal subtraction against these templates provide $\chi^2$ (and $f_\mathrm{star}$) as a function of $T_\mathrm{eff}$. We then fit the minimum of this function in order to estimate both the temperature of the companion and its uncertainties, using the same approach detailed in Section 2.3 of \citet{2025MNRAS.536.2169S}.

We measured temperatures using only the highest signal regions of the spectra -- specifically, in the range 4820--5870\,\AA{}, as indicated in Figure~\ref{fig:optsub}. 
From inferior conjunction ($\phi=0$), we determined an effective temperature of $T_\mathrm{inf} = 5749 \pm 34$\,K, and from superior conjunction ($\phi=0.5$), we determined an effective temperature of $T_\mathrm{sup} = 6106 \pm 35$\,K, or a spectral type ranging from G3 to F9. We also measured temperatures around the first and second points of quadrature (i.e. $\phi=0.25$ and $\phi=0.75$) of $T_\mathrm{q1} = 5868 \pm 37$\,K and $T_\mathrm{q2} = 5879 \pm 39$\,K, respectively, in excellent agreement with each other. 
We found the stellar flux contribution $f_\mathrm{star}$ to cover the range 0.82--0.92, with a maximum value at $\phi = 0$ and a minimum value at $\phi = 0.5$. This indicates a contribution of non-stellar flux or "veiling" of approximately 10--20\% over the full orbit (i.e., $f_\mathrm{veil} \simeq 0.1-0.2$). 
The full optimal subtraction fits and residual spectra for all four phase bins are presented in Figure~\ref{fig:optsubfull}.

\begin{figure}
    \centering
    \includegraphics{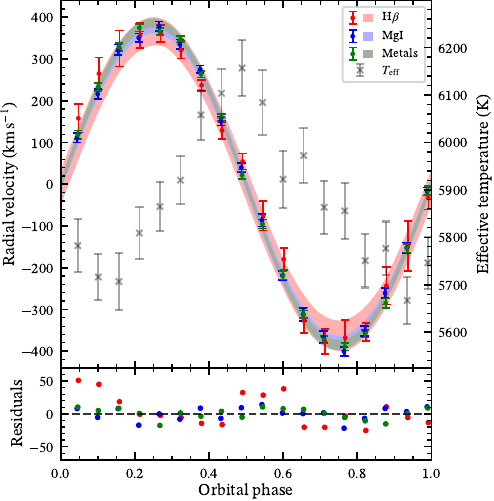}
    \caption{Radial velocity curves of J2055, measured by line-fitting the H$\beta$ line (red), line-fitting the \ion{Mg}{I} triplet (blue), and from cross-correlation using the wide range of metallic line species present over the observed wavelength range (green). Radial velocity fit 1-$\sigma$ errors, as propagated by sampling from MCMC posteriors, are highlighted in matching colours. A temperature curve is also plotted (grey), as measured from optimal subtraction. Residuals to the sinusoidal fits are shown in the lower panel.}
    \label{fig:rvs}
\end{figure}

While $T_\mathrm{inf}$ and $T_\mathrm{sup}$ are in excellent agreement with the photometric temperatures from the simultaneous 2024 data, the increased precision means they are no longer consistent with the photometric temperatures from the 2023 data. This would appear to be indicative of changes in the system between the two epochs, potentially in the strength of the irradiation, or the size and location of hot spots. Additionally, the near-identical temperatures at both points of quadrature would seem to imply J2055's companion lacks asymmetric irradiation. However, looking at the temperatures measured from individual spectra (presented in Figure~\ref{fig:rvs}), it is clear the temperature maximum occurs slightly before phase $0.5$. This suggests the early light curve maximum (Figure~\ref{fig:lcgri}) may be the result of asymmetric irradiation on top of a maximum from ellipsoidal modulation at $\phi = 0.25$. Both the asymmetries present, and the possibility of long-term variability are discussed further in Section~\ref{subsec:discussion5.2}. 

Radial velocities from J2055 were measured in two stages. First, the cross-correlation was computed between the observed spectra and the BT-Settl templates using the full range of absorption line features present, with the exception of the \ion{Na}{I} 5890+5896\,\AA{} doublet (which is contaminated by interstellar absorption). This initial cross-correlation was fit with a sinusoid (with the orbital period fixed to the value from \citetalias{2023ApJ...956..132L}) to obtain a preliminary orbital solution of $K_2 \simeq 382 \kms{}$, $\gamma \simeq 40 \kms{}$, and $T_0 \simeq 60523.987$\,MJD -- where $K_2$ is the radial velocity curve semi-amplitude, $\gamma$ is the systemic velocity, and $T_0$ is the time of companion inferior conjunction. Using this, we applied a careful line-by-line approach (\texttt{astra}'s \texttt{linefitmc} method) to separately fit the H$\beta$ 4861\,\AA{} line and the \ion{Mg}{i} 5167+5172+5183\,\AA{} triplet, in narrow windows around the expected radial velocities. We also computed the cross-correlation against a 5700\,K template (corresponding to the measured $T_\mathrm{inf}$), searching around this orbital solution, over the remainder of the metallic lines in the same wavelengths used for optimal subtraction. Then, using MCMC sampling, as implemented by \texttt{emcee} \citep{emcee}, we fit the measured velocities to determine robust values and errors for $K_2$, $\gamma$, and $T_0$. Thus, we obtained three radial velocity curves from J2055: one from line-fitting of H$\beta$, one from line-fitting the \ion{Mg}{i} triplet, and one from cross-correlation over the other metallic lines in the range 4820--5870\,\AA{}.

The multi-species radial velocity curves from J2055 are presented in Figure~\ref{fig:rvs}. From cross-correlation over the wide range of strong metallic absorption lines (excluding \ion{Mg}{i}), we find a precise radial velocity curve semi-amplitude of $K_\mathrm{2, metals} = 385 \pm 3 \kms{}$. 
From line-fitting of the \ion{Mg}{i} triplet, we measure a slightly lower and less precise (but still consistent) $K_\mathrm{2, \ion{Mg}{i}} = 380 \pm 4 \kms{}$, while from H$\beta$, we find $K_\mathrm{2, H\beta} = 367 \pm 9 \kms{}$. 
The relevant corner plots for all three fits are shown in Figure~\ref{fig:rvcorners}. 
The difference between $K_\mathrm{2, H\beta}$ and $K_\mathrm{2, metals}$ (and also $K_\mathrm{2, \ion{Mg}{i}}$) shows a clear separation between the radial velocities measured from different absorption line species on the surface of J2055's companion. The fact that $K_\mathrm{2, metals} > K_\mathrm{2, H\beta}$ is consistent with Balmer absorption lines being stronger towards the irradiated face of the companion, as has been observed in other spider systems \citep[e.g.][]{2018ApJ...859...54L,2025MNRAS.536.2169S}.

The systemic velocity, $\gamma$, and time of companion inferior conjunction, $T_0$, are consistent between the three fits within their uncertainties. The most precise values are obtained from the cross-correlation results: $\gamma_\mathrm{metals} = 42 \pm 2 \kms{}$ and $T_\mathrm{0, metals} = 60523.9866 \pm 0.0003$\,MJD. 
This $T_0$ differs from a simple projection of \citetalias{2023ApJ...956..132L}'s timing solution to the 2024 observational epoch by almost 7$\sigma$, corresponding to 1 per cent of J2055's orbit. We expect this discrepancy arises from J2055's significant orbital period variations, which can dominate the timing solution's uncertainties over long timescales. Consequently, as mentioned in Section~\ref{sec:optlcJ2055}, we used our own $T_\mathrm{0, metals}$ to phase-fold the photometry rather than $T_\mathrm{asc}$ from \citetalias{2023ApJ...956..132L}. 

For the metal line radial velocities, the reduced $\chi^2$ (i.e. $\chi^2_\nu$) of the fit is slightly over 1: $\chi^2_\mathrm{\nu, metals} = 1.47$, while for the H$\beta$ fits, $\chi^2_\mathrm{\nu, H\beta} = 0.76$. For the \ion{Mg}{i} fit, $\chi^2_\mathrm{\nu, \ion{Mg}{i}} = 1.04$. This suggests that the measurement errors on the H$\beta$ radial velocities are slightly overestimated, while the cross-correlation-based metal radial velocity measurement errors are slightly underestimated. Alternatively, $\chi^2_\mathrm{\nu, metals} > 1$ could imply the metal sinusoidal model fits do not fully capture the data -- considering the effect of irradiation on the centre of light of the absorption features, this is more likely the case. Indeed, the effect of irradiation will distort the observed radial velocity curves, and may be the cause of the apparent trends in the residuals, where the radial velocities depart from a sinusoid. As such, to obtain a more conservative estimate of the parameter errors, the radial velocity errors for each fit have been scaled to obtain $\chi^2_\nu = 1$.

\section{Light curve modelling}
\label{sec:modelling}

We modelled the light curves using the stellar binary light curve synthesis code \textsc{Icarus} \citep{2012ApJ...748..115B}, with the atmosphere grids generated from the \textsc{atlas9} code \citep{1993KurCD..13.....K}. The three photometric datasets, LCO 2023, NOT 2023 and NOT 2024 were converted from AB apparent magnitudes to flux densities, using the zero-point flux density of 3631 Jy for the SDSS \textit{g'}, \textit{r'} and \textit{i'} filters. The orbital phase $0$ was defined as at inferior conjunction of the companion. In order to ensure that each dataset had equal weight, we binned the data so that each dataset had 15 bins. We then fitted these three datasets with a linked sampling algorithm using the nested sampler \textsc{dynesty} \citep{dynesty, dynzen, nested, sampling, bounding}, similar to what was done in \cite{2024ApJ...973..121S}, where we simultaneously fitted all datasets while only allowing some of the parameters to vary independently between the datasets. The rest of the parameters were linked, meaning that they were sampled such that these parameters varied at different steps of the sampler, but were the same for all datasets at each step.

\subsection{Model parameters}
\label{sec:modelparameters}

\textsc{Icarus} requires 11 parameters to be specified for the direct irradiation models, where only the effects of gravity darkening and irradiation contribute to the companion's effective temperature. Four additional parameters are required for hot spot models to specify the size, extent, and location of the spot. We held three of the direct irradiation parameters constant. We assumed a tidally locked system and set the co-rotation factor $\omega = 1$, and fixed the gravity darkening coefficient $\beta = 0.08$ as that corresponds to low-mass companions with a convective envelope \citep{Lucy1967}. From \citetalias{2023ApJ...956..132L}, we fixed the orbital period to $P_\text{orb} = 0.200725452(1) $~days. To model the asymmetry of the light curves, we fitted the data to hot spot models. We determined the location of an equatorial hot spot by finding the orbital phase where the symmetric model under-predicted the flux the most, which was at $\phi = 0.35$. From this, we found the corresponding longitude of the companion hot spot.

\subsubsection{Prior distributions and spectroscopic constraints}

We linked the following parameters, as they are not expected to change on month- to year-timescales: the semi-major axis of the pulsar $x_1$, inclination $i$, projected radial velocity semi-amplitude of the companion $K_2$, filling factor $f$, base temperature $T_\text{base}$, distance $D$, and extinction in the V band $A_V$, which are reported in Table \ref{tab:priors}. From $x_1$ and $P_\text{orb}$, we derived the projected radial velocity semi-amplitude of the pulsar, $K_1$. We used this with $K_2$ to derive the mass ratio $q$ with the constraint from pulsar timing and input it into \textsc{Icarus}. By doing so, we applied a uniform prior on $x_1$ using the reported $x_1 = 0.5996750(1)$~lt-s from \citetalias{2023ApJ...956..132L}. The other parameters were used directly as inputs. The prior on $K_2$ centred around $385 \pm 3 \kms{}$ was derived from spectroscopic analysis, as explained in Section \ref{sec:optrvcJ2055}\footnote{We did not apply any correction to take into account the separation between the companion's centre of light and centre of mass introduced by the light irradiation of this system, as this effect is negligible for the metallic lines (see the detailed discussion in Section \ref{subsec:discussion5.2}).}. Given that edge-on systems are more likely to be detected, we applied a uniform prior on $\cos i$. The prior for the filling factor, defined as $f\equiv r_{\mathrm{nose}}/r_{L_{1}}$, was also uniform between $0$ and $1$. $T_\text{base}$ had a uniform prior between $1000$ and $10000$~K. The other uniform linked prior was $D$, which was between $1$ and $10 \ \mathrm{kpc}$. We assigned to the last linked parameter, $A_V$, a Gaussian prior determined by the colour excess of $E(g-r) = 0.07 \pm 0.01$ from the \cite{2019ApJ...887...93G} dust maps.

The rest of the parameters were unlinked and kept independent. For each dataset, these include the irradiation and hot spot temperatures, $T_\text{irr}$ and $T_\text{spot}$, respectively, as well as the radius of the spot $R_\text{spot}$. For the spot parameters, we applied uniform prior distributions, as seen in Table \ref{tab:priors}. We allowed $T_\text{spot}$ to vary from $0-1000$~K and we allowed the radius to extend up to half of the star. For $T_\text{irr}$, we allow the prior distribution to vary from  $0-10000$~K. We include an additional constraint on the observed $T_\text{inf}$ of the companion for one of our fits. We apply a flat prior of $5746 \pm 165$\,K using the $3 \sigma$ range of the results from optimal subtraction on the single spectrum closest to phase $0$.

\begin{table}
    \centering
    \begin{tabular}{cc}
        \toprule
        Fixed Parameters &  \\
        \midrule
        $P_\text{orb}$ (days) & 0.200725452 \\
        $\omega$ & 1 \\
        $\beta$ & 0.08 \\
        $\theta_\text{spot}$ (deg) & 90 \\
        $\phi_\text{spot}$ (deg) & 306 \\
        \midrule
        Free Parameters & \\
        \midrule
        $x_1$ (lt-s) & $0.5996750 \pm 0.0000001$ \\
        $i$ (deg) & [cos(90), cos(0)] \\
        $K_2$ (\kms{}) & $385 \pm 3$ \\
        $f$ & [0,1] \\
        $T_\text{base}$ (K) & [1000, 10000] \\
        $T_\text{irr}$ (K) & [0, 10000] \\
        $T_\text{inf}$ (K) & [5581, 5911] \\
        $T_\text{spot}$ (K) & [0, 1000] \\
        $R_\text{spot}$ (deg) & [0, 180] \\
        $D$ (kpc) & [1, 10] \\
        $A_V$ (mag) & $0.17 \pm 0.07$\\
        \bottomrule
    \end{tabular}
    \caption{Prior values and distribution of the input parameters. Parameters of the form [minimum, maximum] are uniformly distributed between the values reported. Parameters of the form mean $\pm 1\sigma$ are Gaussian parameters, with the mean and $1\sigma$ uncertainty values reported.}
    \label{tab:priors}
\end{table}

\subsection{Best-fit model results}
\label{sec:bestfitmodelresults}

We find that our two linked models are consistent within $1\sigma$ for all but two of the linked parameters, for the radius of the spot, and for most of the derived parameters, as seen in Table \ref{tab:results}. Among the independent parameters, $R_\text{spot}$ and $T_\text{spot}$ are also compatible within $1\sigma$ between the same epochs of the two fits. The differences in the two fits arise from the independent parameters, which control the effective temperature of the companion. From the fit without any temperature constraints, we find that the model $T_\text{inf}$ is not within $5\sigma$ of the spectroscopic $T_\text{inf}$, depending on the dataset. Similarly for $T_\text{sup}$, the model and spectroscopic temperatures also differ by $5\sigma$. When the $T_\text{inf}$ constraint is implemented during the fitting, we find that the linked $T_\text{base}$ and unlinked $T_\text{irr}$, and in turn the derived $T_\text{inf}$ and $T_\text{sup}$, increase. In addition to this, $A_V$ increases to compensate for the higher temperature, so that the model flux is not over-predicted.

Between the datasets, we find that the hot spot seems to shrink in size from 2023 to 2024, differing more than $1\sigma$ from each other, while the temperature decreases over time. The temperature and radius of the hotspot for the LCO and NOT 2023 datasets are consistent within $1\sigma$ from each other, as is $T_\text{sup}$. On the other hand, we notice that the irradiating temperature $T_\text{irr}$ estimated from LCO 2023 is higher by more than $2-3\sigma$ with respect to both NOT datasets, depending on whether the model included the $T_\text{inf}$ constraint or not. As expected, we see that $T_\text{inf}$ for all epochs for the fit with the $T_\text{inf}$ constraint are consistent to $1\sigma$, but we also note this is the case for the fit without this constraint. The fit without temperature constraints also has $T_\text{sup}$ values that are consistent at less than $1\sigma$, in contrast to the other fit, where the LCO 2023 and NOT 2024 $T_\text{sup}$ values are consistent only at the $1.5\sigma$ level.

\begin{figure*}
    \subfloat{
        \includegraphics[trim={0cm 0cm 1.5cm 0cm}, clip, scale=0.35]{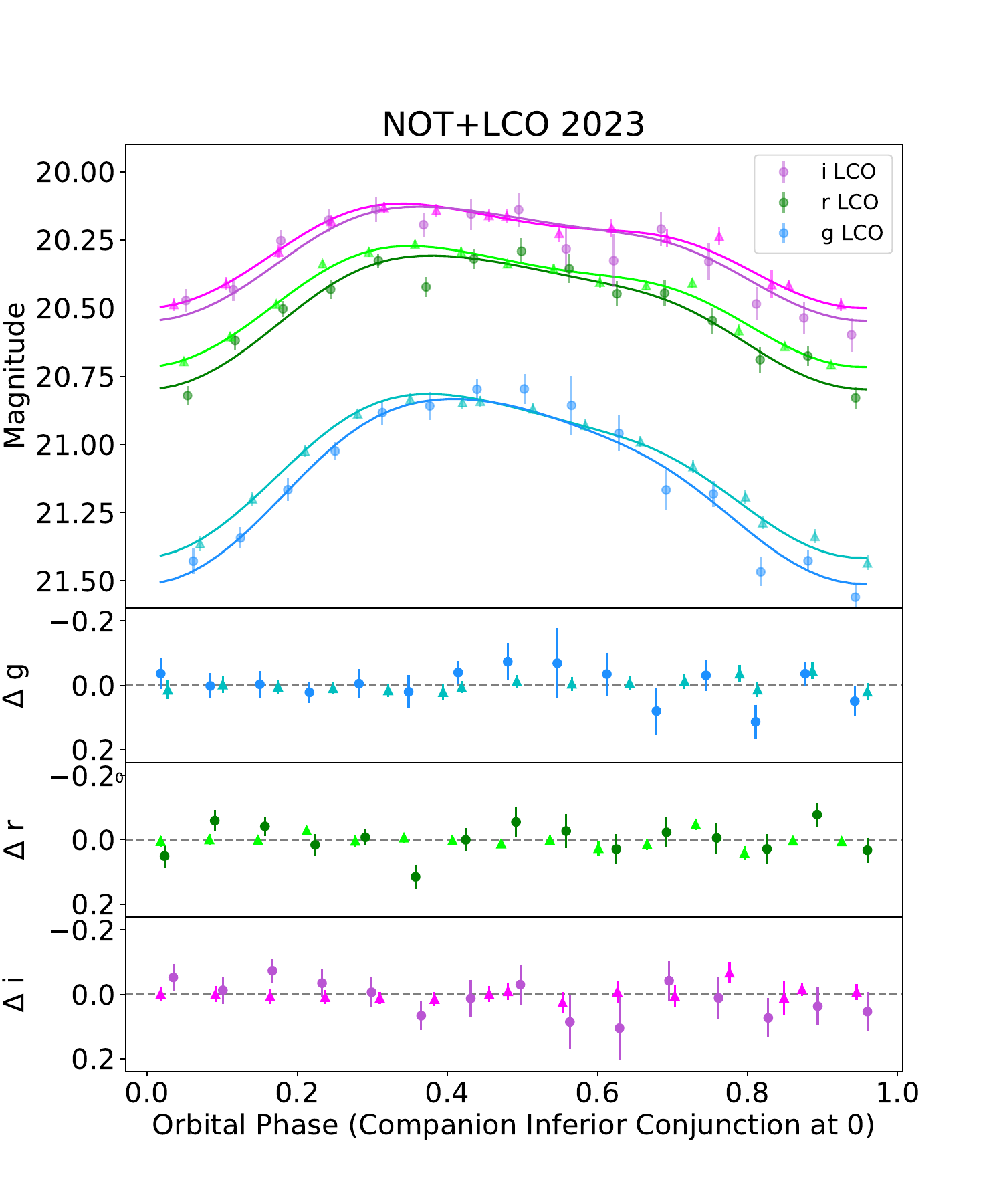}
        }
    \subfloat{
        \includegraphics[trim={1.0cm 0cm 0.5cm 0cm}, clip, scale=0.35]{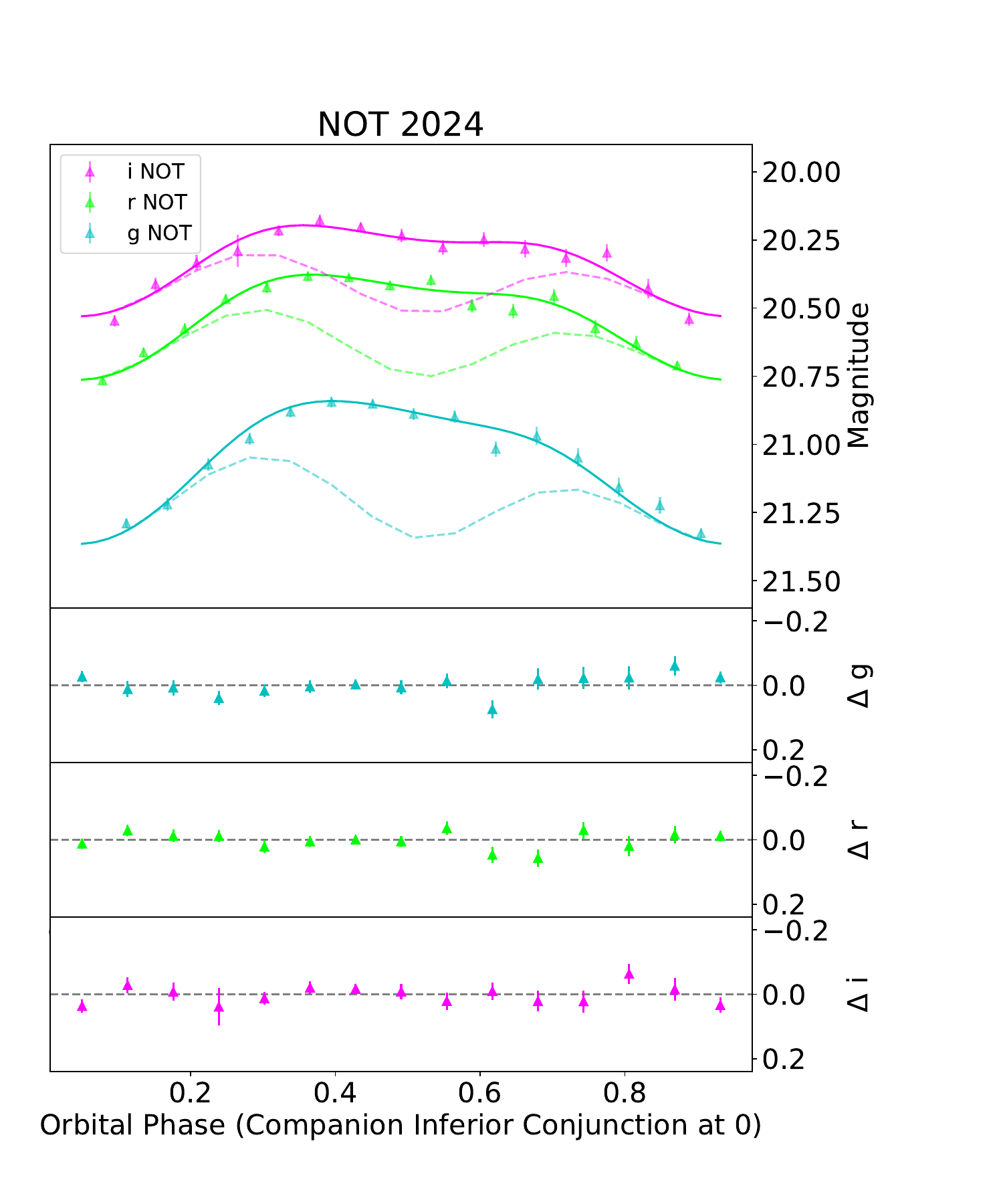}
        }
    \caption{Best-fit model light curves and residuals, with 2023 data plotted on the left and 2024 data plotted on the right. The parameters that were allowed to vary between the three datasets were $T_\text{base}$, $T_\text{irr}$, $T_\text{spot}$ and $R_\text{spot}$. The corner plot for this fit is shown in Figure \ref{fig:lccorner}. Dashed lines on the right panel are the NOT 2024 linked fit model with the irradiation set to 0, so that ellipsoidal modulation is the dominating effect. No other changes were introduced. This effect is explained further in Section \ref{subsec:discussion5.1}.}
    \label{fig:noirrlc}
\end{figure*}

We find that the mass estimates $M_{1}$ and $M_{2}$ for the pulsar and the companion, respectively, are similarly precise for both fits. For the temperature parameters, the $+1\sigma$ relative uncertainties in the fit with the $T_\text{inf}$ constraint are reduced by approximately $1$--$2\%$ compared to the unconstrained fit. While the $A_V$ posterior for the constrained fit does not follow the prior as well as the unconstrained fit and $T_{\mathrm{base}}$ increases in the constrained fit, we find that the other linked parameters are consistent between the two fits. Both the reduced chi-squared values are of $\chi^{2}_{\nu}\simeq1.6$, with the constrained fit more consistent with the spectroscopy given that we use a spectroscopic constraint. Therefore, our best fit is the one with the $T_\text{inf}$ constraint, and our derived best-fit neutron star mass is $1.7^{+0.4}_{-0.1} \ \mathrm{M}_{\sun}$. We show the light curves for this fit in Figure \ref{fig:noirrlc}.

\begin{table*}
    \begin{tabular}{ccccccc}
    \toprule
    Fitted & \multicolumn{3}{c}{no $T$ constraints} & \multicolumn{3}{c}{$T_\text{inf}$ constraint} \\
    & LCO 2023 & NOT 2023 & NOT 2024 & LCO 2023 & NOT 2023 & NOT 2024 \\
    \midrule
    $i$ (deg) & \multicolumn{3}{c}{$76_{-12}^{+9}$}  & \multicolumn{3}{c}{$79_{-13}^{+8}$} \\
    $K_2$ (\kms{}) & \multicolumn{3}{c}{$384_{-2}^{+3}$} & \multicolumn{3}{c}{$384 \pm 2$} \\
    $f$ & \multicolumn{3}{c}{$0.74_{-0.02}^{+0.03}$} & \multicolumn{3}{c}{$0.74_{-0.01}^{+0.03}$} \\
    $T_\text{base}$ (K) & \multicolumn{3}{c}{$5426_{-96}^{+97}$} & \multicolumn{3}{c}{$5728_{-23}^{+40}$} \\
    $T_\text{irr}$ (K) & $4596_{-152}^{+164}$ & $4319_{-137}^{+144}$ & $4288_{-130}^{+141}$ & $4932_{-108}^{+135}$ & $4636_{-71}^{+96}$ & $4606_{-64}^{+103}$ \\
    $T_\text{spot}$ (K) & $340_{-50}^{+61}$ & $334_{-30}^{+37}$ & $230_{-55}^{+99}$ & $383_{-54}^{+67}$ & $373_{-28}^{+41}$ & $265_{-67}^{+141}$ \\
    $R_\text{spot}$ (deg) & $49 \pm 8$ & $55_{-7}^{+9}$ & $34_{-9}^{+14}$ & $49_{-8}^{+9}$ & $54_{-6}^{+7}$ & $33_{-10}^{+16}$ \\
    $D$ (kpc) & \multicolumn{3}{c}{$4.93_{-0.29}^{+0.54}$} & \multicolumn{3}{c}{$5.34_{-0.20}^{+0.54}$} \\
    $A_V$ (mag) & \multicolumn{3}{c}{$0.20_{-0.06}^{+0.08}$} & \multicolumn{3}{c}{$0.32 \pm 0.05$} \\
    \midrule
    Derived & & & & & & \\
    \midrule
    q & \multicolumn{3}{c}{$5.90 \pm 0.04$} & \multicolumn{3}{c}{$5.90_{-0.03}^{+0.04}$} \\
    $M_1 (\mathrm{M}_{\sun})$ & \multicolumn{3}{c}{$1.8_{-0.1}^{+0.4}$} & \multicolumn{3}{c}{$1.7_{-0.1}^{+0.4}$} \\
    $M_2 (\mathrm{M}_{\sun})$ & \multicolumn{3}{c}{$0.30_{-0.02}^{+0.07}$} & \multicolumn{3}{c}{$0.29_{-0.01}^{+0.07}$} \\
    $T_\text{sup}$ (K) & $5887_{-162}^{+139}$ & $5816_{-160}^{+137}$ & $5713_{-157}^{+135}$ & $6255_{-144}^{+118}$ & $6172_{-142}^{+117}$ & $6062_{-140}^{+115}$ \\
    $T_\text{inf}$ (K) & $5346_{-147}^{+127}$ & $5363_{-147}^{+127}$ & $5314_{-146}^{+126}$ & $5643_{-130}^{+107}$ & $5657_{-131}^{+107}$ & $5607_{-106}^{+129}$ \\
    $L_\text{irr}$ ($10^{32} \ \mathrm{erg} \ \mathrm{s}^{-1}$) & $1.8 \pm 0.7$ & $1.4 \pm 0.6$ & $1.3 \pm 0.5$ & $2.3 \pm 0.9$ & $1.8 \pm 0.7$ & $1.8 \pm 0.7$ \\
    \midrule
    Model Fit & & & & & & \\
    \midrule
    $\chi^2_\nu$ & \multicolumn{3}{c}{1.59} & \multicolumn{3}{c}{1.66} \\
    \textit{g'} Offset & 0.085 & 0.008 & -0.023 & 0.121 & 0.040 & -0.044 \\
    \textit{r'} Offset & 0.036 & 0.032 & -0.047 & 0.015 & -0.056 & -0.083 \\
    \textit{i'} Offset & -0.008 & -0.043 & -0.095 & -0.046 & -0.083 & -0.056 \\
    \bottomrule
    \end{tabular}
    \caption{Best-fit linked model results, with the 50th percentile value reported with the 16 and 84 percentiles as uncertainties. The fit without any spectroscopic temperature constraints is on the left and the fit with the constraint on $T_\text{night}$ is on the right. For the independent parameters, the ordering of the results by datasets is as follows: LCO 2023, NOT 2023, and NOT 2024. Fitted parameters are reported at the top of the table and derived parameters in the middle. The mass ratio $q$ reported is of the form $M_{1}/M_{2}$. Model reduced $\chi^2$ and band offsets are reported on the bottom third.}
    \label{tab:results}
\end{table*}


\section{Discussion}
\label{sec:discussion}

\subsection{A new mildly-irradiated redback}
\label{subsec:discussion5.1}

The optical light curves of J2055 do not indicate any state change between July--November 2023 and August 2024, as we can see in Figure \ref{fig:lcgri}. In both epochs, this system consistently shows variable colours and a single asymmetric broad light maximum per orbit at phase $\simeq0.4$, slightly anticipating companion superior conjunction at phase $0.5$. This strongly suggests that the companion star of J2055 is heated by the pulsar wind with an asymmetric pattern. This is further supported by the effective temperature curve shown in Figure \ref{fig:rvs}, which peaks slightly before phase $0.5$.

Furthermore, we measure peak-to-peak amplitudes of $0.6$, $0.5$ and $0.4 \ \mathrm{mag}$ in \textit{g'}, \textit{r'} and \textit{i'}, respectively, which are lower than the typical amplitudes of $\gtrsim 1 \ \mathrm{mag}$ observed from irradiated redbacks (see e.g. \citealt{2018ApJ...859...54L} for PSR~J2215$+$5135 and \citealt{2011ApJ...743L..26R} for PSR~J2339$-$0533). We also observe a small hump across the companion descending node (phase $0.75$), which appears sharper in the \textit{r'} and \textit{i'} bands compared to \textit{g'}. This is likely produced by an underlying ellipsoidal modulation, still noticeable in the redder bands but overwhelmed in \textit{g'} where the irradiation is stronger, and is confirmed by the model without irradiation shown in the right panel of Figure \ref{fig:noirrlc}. All these features indicate that the companion star of J2055 is mildly heated by the pulsar wind, lying in-between the ellipsoidal and the strong irradiation regimes \citep{2023MNRAS.525.2565T}.  

A few other RB systems show optical phenomenology similar to that of J2055. In particular, PSR~J1048$+$2339 (henceforth called J1048) has been observed to change from an ellipsoidal to an irradiated state in less than two weeks \citep{2019A&A...621L...9Y}.
The irradiation-dominated optical light curves of J1048 show the same shape as J2055's light curves, with asymmetric flux maxima and small humps at about phase $0.75$. Interestingly, J2055 seems to be much more stable than J1048, without any clear state transition during our 4-month-long LCO monitoring. Among other RBs, PSR~J1306$-$40 \citep[J1306,][]{2019ApJ...876....8S} and the two transitional MSPs PSR~J1023$+$0038 and PSR~J1227$-$4853 (J1023 and J1227, respectively, \citealt{2021MNRAS.507.2174S}\footnote{\cite{2021MNRAS.507.2174S} carried out optical photometry of the two transitional MSPs J1023 and J1227 during their rotation-powered state.}) exhibit optical light curves closely resembling those of J2055. All show amplitudes of $\lesssim 0.8 \mathrm{mag}$ and asymmetric flux maxima flattening at redder bands, where the ellipsoidal variation contribution is non-negligible compared to the irradiation component. 

To quantify the irradiation power of spider MSPs we can apply the pulsar spin-down to companion flux ratio parameter $f_{\mathrm{sd}}$ (see Equation 3 in \citealt{2023MNRAS.525.2565T}). Concerning J2055, using $L_{\mathrm{sd}}=(3.83\pm0.01) \times10^{34} \ \mathrm{erg} \ \mathrm{s}^{-1}$ and $P_{\mathrm{orb}}=4.81741085\pm0.00000002 \ \mathrm{h}$ from \citetalias{2023ApJ...956..132L}, $T_{\mathrm{base}}=5728^{+40}_{-23} \ \mathrm{K}$
and $M_{1}+M_{2}=2.0\pm0.2\ \mathrm{M}_{\sun}$ from our light curve modelling, we obtain a flux ratio of $f_{\mathrm{sd}}=3.0\pm0.1$, placed exactly in the transition region between ellipsoidal and irradiated regimes \citep{2023MNRAS.525.2565T}, as we expect. We also find $f_{\mathrm{sd}}=2.6\pm0.2$, $3.9\pm0.4$ and $4.9\pm1.2$ for J1048, J1023 and J1227\footnote{We could not compute the flux ratio parameter for J1306, as this system does not have any spin-down luminosity estimate nor radio timing solution available in the literature.}, respectively \citep{2023MNRAS.525.2565T} -- all consistent with the same intermediate state observed in J2055.

Furthermore, our best-fit model of J2055's optical light curves yields an irradiating luminosity of $L_{\mathrm{irr}}=(1.8\pm0.7) \times 10^{32} \ \mathrm{erg} \ \mathrm{s}^{-1}$, which remains consistent across the three datasets -- LCO 2023, NOT 2023, and NOT 2024. Following the definition by \cite{2013ApJ...769..108B}, we derive an irradiation efficiency relative to the spin-down luminosity of $\epsilon_{\mathrm{irr}}\equiv L_{\mathrm{irr}}/L_{\mathrm{sd}}=0.5\pm0.2 \%$ for this system. In comparison, J2055 irradiation efficiency is lower than the $10$--$30\%$ range observed in strongly irradiated spider companions \citep{2013ApJ...769..108B}, yet higher than the $\simeq0.1\%$ of PSR~J1622$-$0315 \citep{2024ApJ...973..121S}, whose light curves exhibit little to no irradiation. This further supports the conclusion that J2055’s companion is mildly irradiated by the pulsar wind.

The close resemblance of the optical light curves of J2055, J1048 and J1306 with that of J1023 and J1227 suggests that they might all be transitional MSPs observed during their rotation-powered state. 
We argue that these systems may show in the future a transition from the radio pulsar to the accretion disc state. Indeed, the lack of strong irradiation indicates that the magnetic pressure exerted by the pulsar wind may not be effective enough to overcome the formation of an accretion disc in these systems, in case the companion star exceeds its Roche lobe. Although we estimate for J2055 a filling factor $f\equiv r_{\mathrm{nose}}/r_{L_{1}}$ of only $0.74^{+0.03}_{-0.01}$ (see Table \ref{tab:results}), the other RBs considered are also significantly under-filling their Roche lobes \citep{2019A&A...621L...9Y,2019ApJ...876....8S,2021MNRAS.507.2174S}. In particular, \cite{2021MNRAS.507.2174S} estimate $f=0.94$ and $0.84$ for J1023 and J1227, respectively, using the hot spot model. However, their corresponding volume-averaged filling factors are much closer to unity
due to tidal distortion \citep{2021MNRAS.507.2174S}, namely $f_{\mathrm{VA}}=0.99$ and $0.96$ for J1023 and J1227, respectively (see Figure 10 in \citealt{2021MNRAS.507.2174S}). Using \textsc{Icarus}, we convert J2055's filling factor into a volume-averaged filling factor of $f_{\mathrm{VA}}=0.90\pm0.01$, which suggests a bloated companion star that may unpredictably overfill its Roche lobe and start transferring mass.
\subsection{Radial velocities and spectroscopic temperatures}
\label{subsec:discussion5.2}

The optical spectra of J2055 enabled precise measurements of the companion's temperature throughout its orbit, with the results of optimal subtraction yielding $T_{\mathrm{inf}}=5749\pm34 \, \mathrm{K}$ and $T_{\mathrm{sup}}=6106\pm35 \, \mathrm{K}$, respectively (Section \ref{sec:optrvcJ2055} and Figure \ref{fig:optsubfull}). These values are in full agreement with the less precise estimates derived from the simultaneous 2024 photometry, while they are both higher than those from the 2023 data of $T_{\mathrm{inf}}=5300\pm200 \, \mathrm{K}$ and $T_{\mathrm{sup}}=5900\pm200 \, \mathrm{K}$.

Since the temperatures measured around inferior and superior conjunction depend on both the binary inclination and the irradiating flux \citep{2025MNRAS.536.2169S}, the observed differences between 2023 and 2024 could suggest an increase in the irradiation strength of J2055. However, the light curve modelling presented in Section \ref{sec:modelling} indicates that $T_{\mathrm{inf}}$ and $T_{\mathrm{sup}}$ remain consistent across the two epochs, instead revealing a change in the size of the companion's hot spot (see Table \ref{tab:results}). To better constrain the inclination and base temperature ($T_\mathrm{base}$) of this system, we incorporated the spectroscopic temperature at inferior conjunction ($T_\mathrm{inf}$) as a prior in one of our linked fits, as detailed in Section \ref{sec:modelparameters}.

We also estimated spectroscopic temperatures at the companion's ascending and descending nodes for this system, obtaining $T_\mathrm{q1} = 5868 \pm 37$\,K and $T_\mathrm{q2} = 5879 \pm 39$\,K, respectively. Their close agreement, along with the early occurrence of the light curve and temperature maxima (Figures \ref{fig:lcgri} and \ref{fig:rvs}), suggests the presence of a light asymmetric irradiation component superimposed on an ellipsoidal modulation peak at the first quadrature. This interpretation is further supported by the light curve model without irradiation, shown in the right panel of Figure \ref{fig:noirrlc} with dashed lines, which is only slightly exceeded by the asymmetric irradiation model (solid lines) at phases $0.25$ and $0.75$. Future observations of J2055 in a potential ellipsoidal-dominated state, similar to the scenario analysed by \citet{2019A&A...621L...9Y} for J1048, could help disentangle and quantify these two different contributions.

As we can see from Figure \ref{fig:optsub}, J2055's spectra show numerous metallic absorption lines, which appear most prominent around phase $0$, when the cold side of the companion is visible. In contrast, the H$\beta$ line becomes stronger around the companion superior conjunction, when we observe the companion's hot side. As a result, the semi-amplitude of the radial velocity curve measured from the H$\beta$ line ($K_\mathrm{2, H\beta} = 367 \pm 9 \kms{}$) is lower than that derived from the metallic lines ($K_\mathrm{2, metals} = 385 \pm 3 \kms{}$). This discrepancy is due to the displacement of the centre of light relative to the companion’s centre of mass, with stronger irradiation causing a more pronounced shift of the Balmer lines toward the inner surface of the companion.

The separation between the centre of light and the CoM of the companion has already been observed in a few other irradiated spiders (e.g. PSR~J2215$+$5135, \citealt{2018ApJ...859...54L}; PSR~J1810$+$1744, \citealt{2021ApJ...908L..46R} and J1048, \citealt{2025MNRAS.536.2169S}). Here, we did not apply any `$K$-correction' to take this effect into account. While metallic lines appear stronger on the night side of the companion, irradiation causes the continuum flux to be higher towards the day side. The combination of these two factors can result in metallic lines instead tracking the CoM reasonably well, while Balmer lines are pushed even more towards the inner face \citep{2024MNRAS.528.4337D}. Considering this, and the fact that J2055's companion is only lightly irradiated, the metallic line RVs were taken to be approximately representative of the movement of the companion's CoM, and thus $K_\mathrm{2, metals}$ was used as a prior on the CoM $K_2$ for the light curve modelling (see Table \ref{tab:priors}).

The optical spectra of J2055 exhibit no emission features, ruling out the presence of an accretion disc in the system, as expected. If an accretion disc were present, we would expect to detect prominent double-peaked hydrogen or helium emission lines, similar to those observed in the transitional millisecond pulsar J1023 during its disc state \citep{2019MNRAS.488..198S}. Interestingly, although the optical light curves of J2055 closely resemble those of J1048, the latter displays strong H$\alpha$ emission lines in its spectra \citep{2021A&A...649A.120M,2025MNRAS.536.2169S}. This emission has been interpreted by \cite{2021A&A...649A.120M} as an effect of the intrabinary shock between the pulsar wind and the material ablated in close proximity of the companion star. However, it is important to note that such features have only been observed in a few spider pulsars under poorly understood conditions \citep{2019ApJ...872...42S}. Therefore, the absence of H$\alpha$ emission in J2055 does not necessarily imply the lack of an intrabinary shock in this system.

\subsection{Masses, orbital parameters and distance}
\label{subsec:discussion5.3}

We modelled the optical light curves of J2055, incorporating an equatorial hot spot component to account for the asymmetry observed in their maxima, as the symmetric irradiation model showed significant trends in the residuals. In our modelling, parameters expected to remain stable across the three datasets -- LCO 2023, NOT 2023, and NOT 2024 -- were linked (see Section \ref{sec:modelparameters} for details). To reduce the model degeneracy between the orbital inclination and the companion’s radial velocity semi-amplitude, we adopted a prior of $K_\mathrm{2} = 385 \pm 3 \kms{}$ based on our spectroscopic analysis. This is crucial for accurately determining the neutron star mass, which scales as $M_{1}\propto K_{2}^{3}/\sin^{3}{i}$.

We also incorporated an additional spectroscopic constraint on the observed temperature at inferior conjunction, $T_\mathrm{inf}$, to obtain more accurate estimates of the base temperature of the companion, the strength of irradiation, and the orbital inclination from light curve modelling. Temperature measurements obtained using optimal subtraction (Section \ref{sec:optrvcJ2055}) rely on absorption line strengths, making them independent of extinction and therefore more robust than those inferred from optical light curves alone. Both models, with and without the $T_{\mathrm{inf}}$ constraint, provide good fits to J2055's light curves, yielding similar reduced $\chi^{2}$ values of $1.59$ and $1.66$, respectively. However, the $T_{\mathrm{inf}}$ derived from the `unconstrained' fit is significantly lower than the spectroscopic estimate, and the uncertainties for the `constrained' fit are smaller overall, as we can see in Table \ref{tab:results}. Therefore, we adopt the fit with the $T_{\mathrm{inf}}$ constraint as our best model, with the corresponding system parameters reported in the right section of Table \ref{tab:results}.

Our best-fit model shows that both the observed temperatures and the irradiating luminosity remain consistent across the three epochs, with values of $T_{\mathrm{inf}} \simeq 5600 \, \mathrm{K}$, $T_{\mathrm{sup}} \simeq 6200 \, \mathrm{K}$, and $L_{\mathrm{irr}} \simeq 2 \times 10^{32} \, \mathrm{erg} \, \mathrm{s}^{-1}$, respectively. On the other hand, the companion hot spot in LCO and NOT 2023 data is both slightly warmer and slightly larger than the hot spot observed in the NOT 2024 dataset. Comparing NOT 2023 to NOT 2024, the hot spot shrinks from $54^{\circ}$ to $33^{\circ}$ and cools by $\sim$$100\,\mathrm{K}$. We propose that this hot spot is produced by variable asymmetric irradiation from an intrabinary shock between the pulsar and companion winds. The location of the spot on the companion (see Table \ref{tab:priors}) can be explained by assuming that the shock is wrapped around the pulsar, as has been observed in other RB systems \citep{2018ApJ...866...71C}, with the companion wind dominating over the pulsar wind \citep{2016ApJ...828....7R}. This shock geometry would enhance the irradiating flux on the companion's trailing edge, which accounts for the early light curve maximum observed in J2055.

Here, we present the first estimates for J2055's orbital inclination, $i=79^{+8}_{-13}$$^\circ$, and neutron star mass, $M_{1}=1.7^{+0.4}_{-0.1} \ \mathrm{M}_{\sun}$, from our best-fit model with the $T_{\mathrm{inf}}$ constraint, indicating a fairly massive neutron star\footnote{See Figure 3 in \citealt{2020mbhe.confE..23L} for some of the most massive neutron stars.} hosted in a moderately edge-on system. In general, the precision of neutron star mass measurements in spiders largely depends on the orbital inclination and the companion's radial velocity semi-amplitude. As we can see in Figures \ref{fig:lccornernotn} and \ref{fig:lccorner}, the area covered by $M_{1}$ in both the unconstrained and constrained fits is determined by $i$, while it shows no correlation with $K_{2}$, as the latter is tightly inferred from spectroscopy.
This leads to a more accurate estimate of $T_{\mathrm{base}}$, $L_{\mathrm{irr}}$, and consequently $i$, as the observed temperatures and light curve modulation are tightly linked to these parameters. In this case, where the companion radial velocity is precisely determined, the uncertainty on the orbital inclination drives the precision of the neutron star mass measurement.

Among other parameters, we obtain a companion mass of $M_{2}=0.29^{+0.07}_{-0.01} \ \mathrm{M}_{\sun}$, in full agreement with the median value $0.29 \ \mathrm{M}_{\sun}$ inferred by \citetalias{2023ApJ...956..132L}. Interestingly, the distance from our best fit $D=5.34^{+0.54}_{-0.20} \ \mathrm{kpc}$ places J2055 farther than the estimates from \citetalias{2023ApJ...956..132L}, which found distances of $2.4 \ \mathrm{kpc}$ and $3.7 \ \mathrm{kpc}$ using the electron density models of \cite{2017ApJ...835...29Y} and \cite{2002astro.ph..7156C}, respectively. This discrepancy is not surprising, as dispersion measure distances for spider MSPs are known to systematically underestimate the true distances due to small-scale inaccuracies in the electron density models, as shown by \cite{2024MNRAS.529..575K}.

Spider MSPs often emit X-rays through synchrotron radiation from the intrabinary shock region \citep{2014ApJ...783...69G,2014ApJ...795...72L}. Using our updated $D=5.34$~kpc, we revise the upper limit on the X-ray luminosity of J2055 (previously estimated by \citetalias{2023ApJ...956..132L})
to $L_X < 1.9\times10^{31} \ \mathrm{erg} \ \mathrm{s}^{-1}$ (0.3--10 keV). This new value is more consistent with the typical X-ray luminosities of Galactic RBs, whose distribution peaks at $8\times10^{31} \ \mathrm{erg} \ \mathrm{s}^{-1}$ \citep{2023MNRAS.525.3963K}. Likewise, assuming the association of this RB MSP with 4FGL~J2055.8$+$1545, we revise its $\gamma$-ray luminosity
to $L_\gamma = 7.8\times10^{33} \ \mathrm{erg} \ \mathrm{s}^{-1}$ (0.1--100 GeV), which remains compatible with the typical $\gamma$-ray luminosity ranges of MSPs \citep{2023ApJ...958..191S}.

\section{Conclusions}
\label{sec:6}

We obtained the first optical light and radial velocity curves of the companion to the RB MSP J2055. The light curve exhibits a mildly-irradiated regime with amplitudes of $\simeq0.4$--$0.6 \ \mathrm{mag}$, where the heating from the pulsar wind slightly surpasses the underlying ellipsoidal modulation. The light maximum shows a pronounced asymmetry, occurring earlier than the superior conjunction. Through optimal subtraction of the companion's optical spectra, we accurately determine temperatures over its orbit, which show an amplitude of about $400 \, \mathrm{K}$.

Using independent constraints on the companion's radial velocity semi-amplitude and temperature from our spectroscopy, we simultaneously modelled three optical light curves of J2055 obtained at different epochs, linking the parameters $T_{\mathrm{base}}$, $L_{\mathrm{irr}}$, $T_{\mathrm{spot}}$ and $R_{\mathrm{spot}}$. While these parameters remained largely stable from July 2023 to August 2024, the hot spot component exhibited a slight variation between epochs, which we attribute to an asymmetric intrabinary shock. Our analysis provides, among others, the first measurements of the orbital inclination, $i=79^{+8}_{-13}$$^\circ$, and neutron star mass, $M_{1}=1.7^{+0.4}_{-0.1} \ \mathrm{M}_{\sun}$, of J2055, favouring a moderately inclined system hosting a fairly massive neutron star. Additionally, the inferred distance of $5.3 \ \mathrm{kpc}$ places J2055 farther than the dispersion measure estimates of $2.4$--$3.7 \ \mathrm{kpc}$ derived from radio timing of the MSP.

The distinctive shape of J2055's optical light curve and its mild level of irradiation closely resemble those observed in the transitional MSPs PSR~J1023$+$0038 and PSR~J1227$-$4853 during their rotation-powered states. These similarities, combined with a Roche-lobe filling factor comparable to that of the two transitional systems, suggest that J2055 could potentially transition to a disc accretion-powered state in the future.

\section*{Acknowledgements}
This project has received funding from the European Research Council
(ERC) under the European Union’s Horizon 2020 research and innovation
programme (grant agreement No. 101002352).
This work makes use of observations from the Las Cumbres Observatory global telescope network (we thank L. Storrie-Lombardi and N. Volgenau for their support).
Based on observations made with the Nordic Optical Telescope, operated jointly by Aarhus University, the University of Turku and the University of Oslo, representing Denmark, Finland and Norway, the University of Iceland and Stockholm University.
Based on observations made with the GTC telescope, in the Spanish
Observatorio del Roque de los Muchachos of the Instituto de
Astrof{\'i}sica de Canarias, under Director’s Discretionary Time
(GTC2024-249).
We thank T. Marsh for the use of {\sc molly} and {\sc pamela}. MT thanks K. Koljonen for a discussion on the irradiation strength in spider systems, and how it affects the shape of their optical light curves.
JC acknowledges support by the Spanish Ministry of Science via the Plan de Generaci\'on de Conocimiento through grant PID2022-143331NB-100.
RPB is supported by the UK Science and Technology Facilities Council (STFC), grant number ST/X001229/1.
\section*{Data availability}
The raw LCO and NOT images with bias and flats frames used for data reduction can be obtained by contacting M. Turchetta. The raw GTC science and arc spectra with corresponding data reduction frames can be obtained by contacting J. Simpson.




\bibliographystyle{mnras}
\bibliography{main}{} 




\appendix
    \label{fig:LCOmultiPDM}

\section{Details of spectral analysis}
\label{app:optsub}

\begin{figure*}
    \centering
    \includegraphics{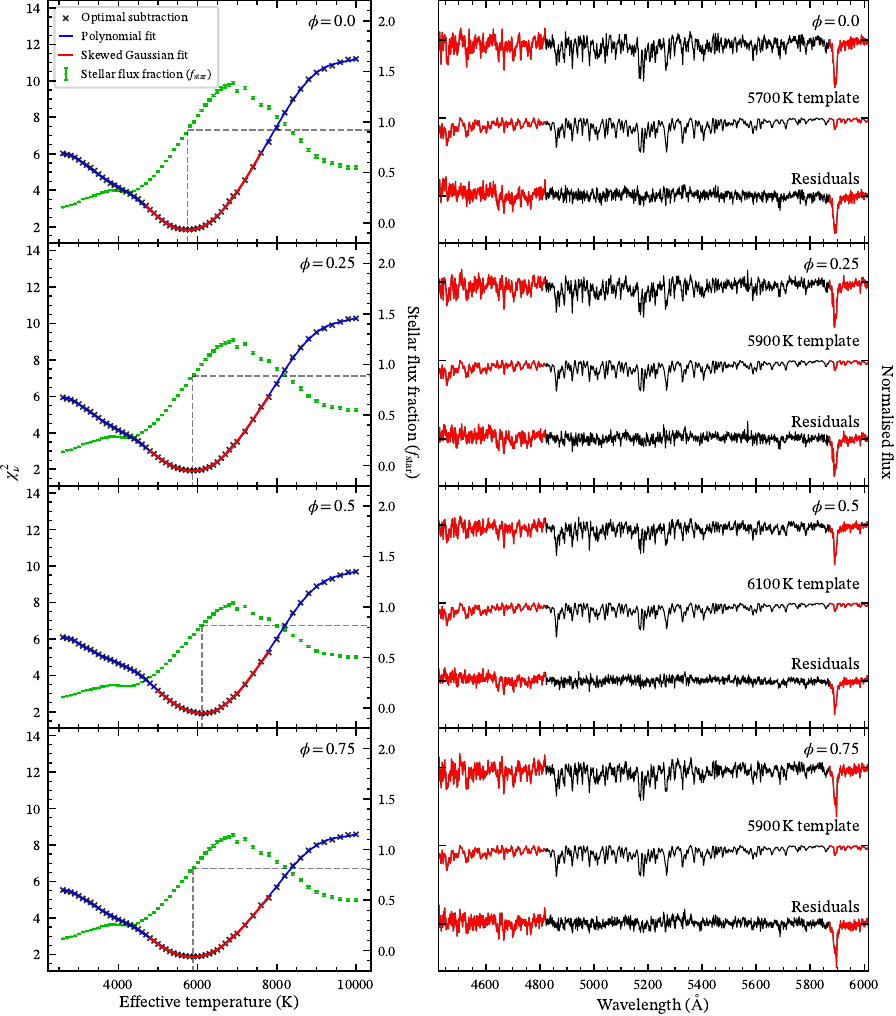}
    \caption{Optimal subtraction results over the full orbit of J2055. Results from average spectra, averaged over bins of a quarter phase in width, centred around companion inferior conjunction ($\phi=0$), first point of quadrature ($\phi=0.25$), companion superior conjunction ($\phi=0.5$), and second point of quadrature ($\phi=0.75$), from top to bottom.\\
    The left panel shows the raw optimal subtraction results for each average spectrum: reduced $\chi^2$ ($\chi^2_\nu$, black) as a function of template temperature, alongside the stellar flux fraction $f_\mathrm{star}$ (i.e. the optimal factor scaling the templates) in green. In blue, the global polynomial fit is shown, while the fit to the minimum (a skewed Gaussian function) is shown in red. Dashed lines indicate the best-fit temperatures and their corresponding $f_\mathrm{star}$ values.\\
    The right panel shows, at each of the four phases, the full, average spectrum (top), the closest matching template (middle), and the optimal subtraction residuals (bottom). For all three spectra, the wavelength ranges used for optimal subtraction are shown in black, while excluded ranges are highlighted in red.}
    \label{fig:optsubfull}
\end{figure*}

\begin{figure*}
    \centering
    \includegraphics{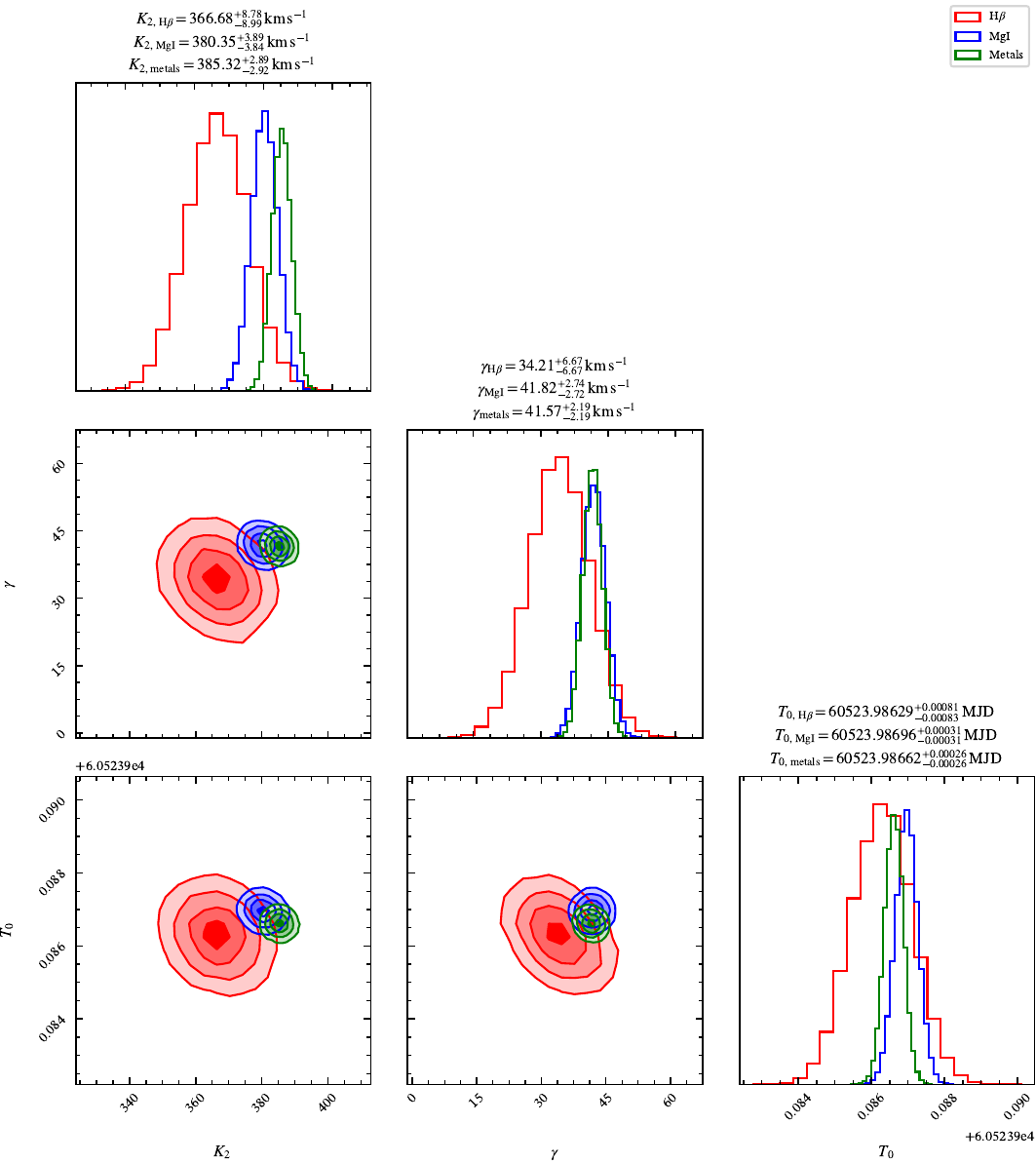}
    \caption{Corner plots for the radial velocity curves presented in Figure~\ref{fig:rvs}. Following the same colour scheme, red, blue, and green histograms and contours represent sinusoidal fits to radial velocities measured by line-fitting of H$\beta$, line-fitting the \ion{Mg}{I} triplet, and cross-correlation over the wide range of metallic line species present in the spectra, respectively.}
    \label{fig:rvcorners}
\end{figure*}

\section{Light Curve Modelling Corner Plot}
\begin{figure*}
    \centering\includegraphics[scale=0.18]{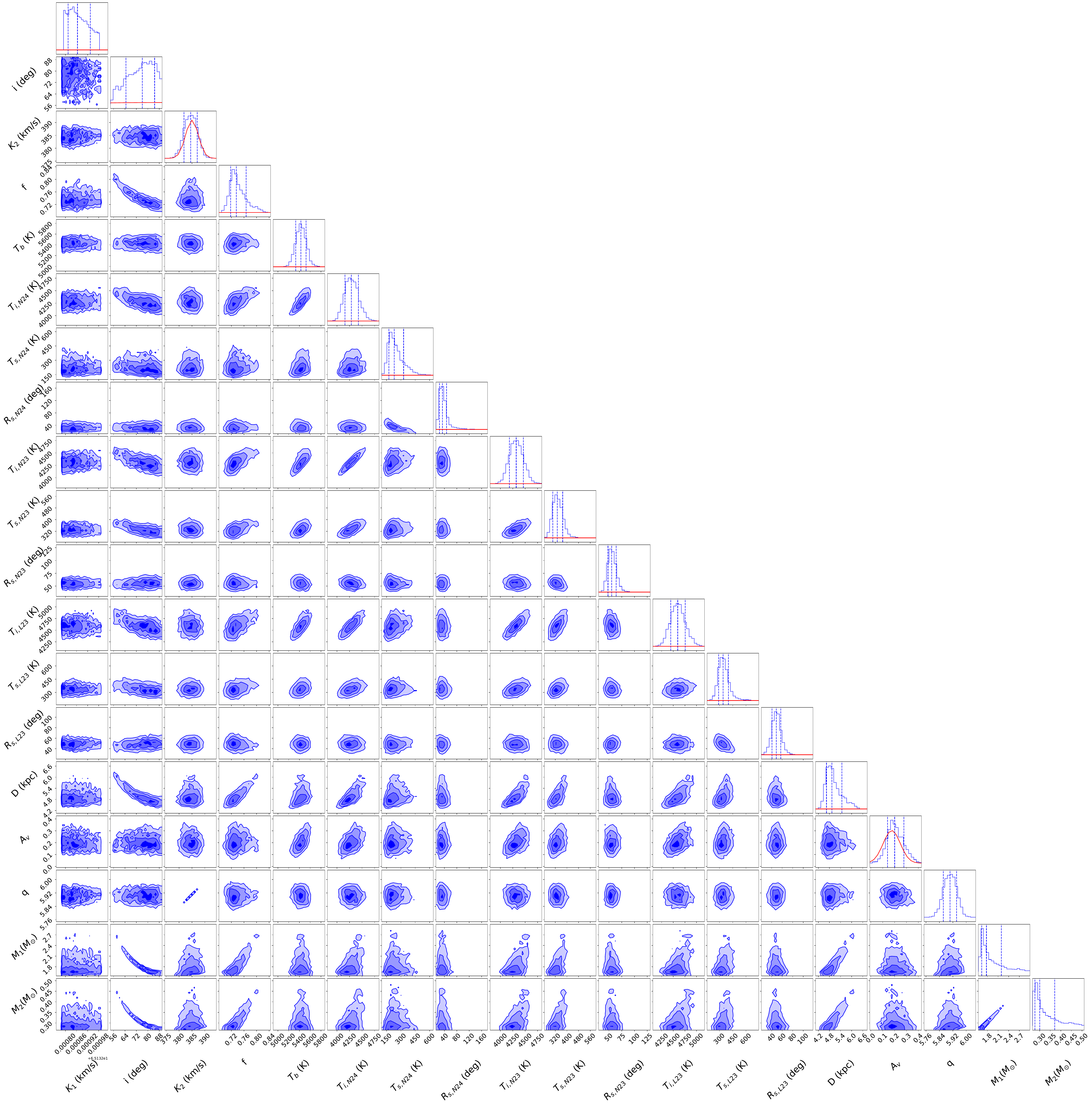}
    \caption{Corner plot of the linked fit with hot spots for the three datasets without constraints on $T_\text{inf}$. The parameters $T_{\mathrm{base}}$, $T_{\mathrm{irr}}$, $T_{\mathrm{spot}}$ and $R_{\mathrm{spot}}$ have been abbreviated as $T_{\mathrm{b}}$, $T_{\mathrm{i}}$, $T_{\mathrm{s}}$ and $R_{\mathrm{s}}$, respectively, to prevent overlap of the axis labels. Additionally, the LCO 2023, NOT 2023, and NOT 2024 datasets are denoted by the suffixes `L23', `N23', and `N24', respectively.}
    \label{fig:lccornernotn}
\end{figure*}

\begin{figure*}
    \centering\includegraphics[scale=0.18]{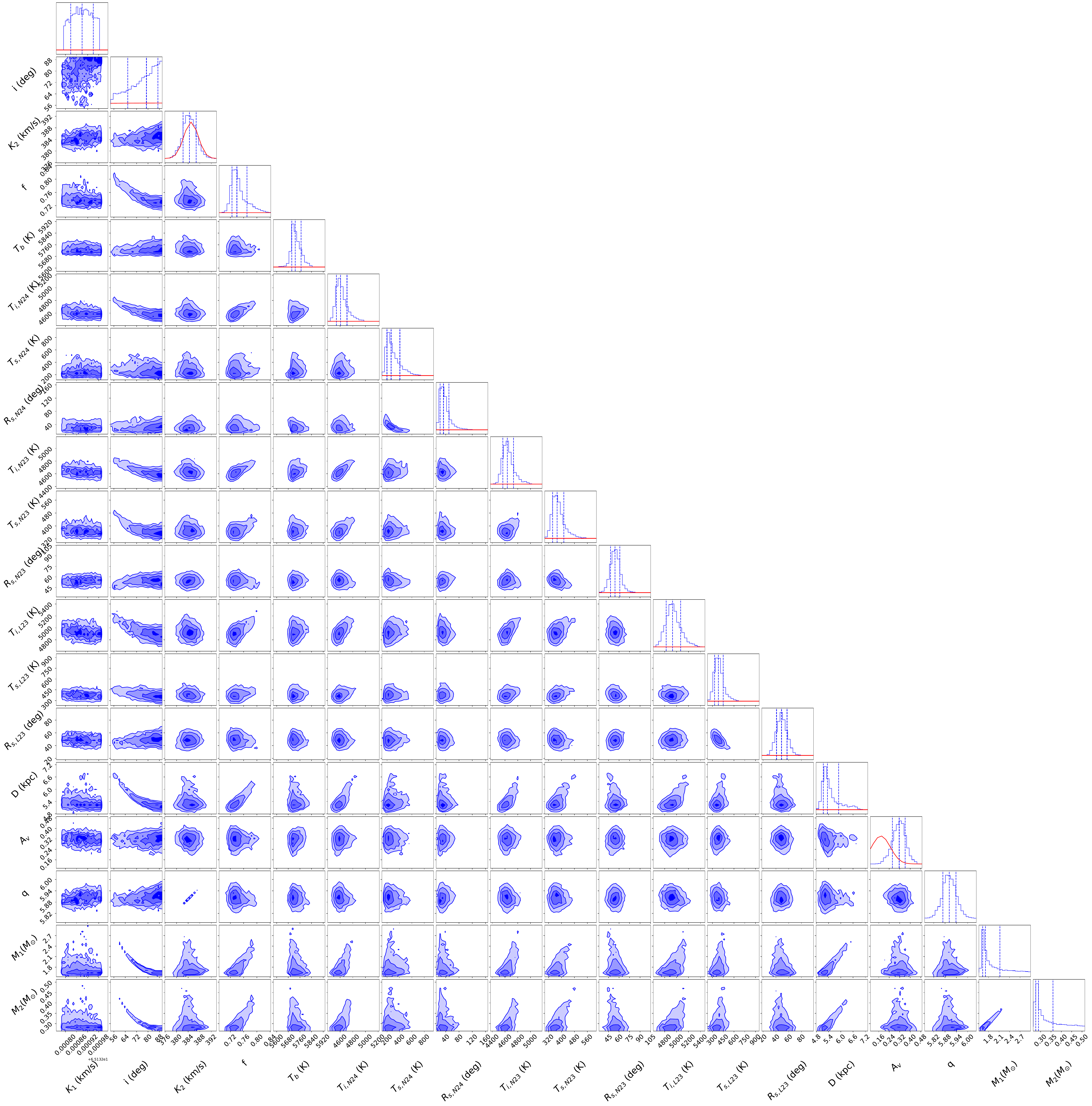}
    \caption{Corner plot of the linked fit with hot spots for the three datasets with the spectroscopic constraint on $T_\text{inf}$. The parameters $T_{\mathrm{base}}$, $T_{\mathrm{irr}}$, $T_{\mathrm{spot}}$ and $R_{\mathrm{spot}}$ have been abbreviated as $T_{\mathrm{b}}$, $T_{\mathrm{i}}$, $T_{\mathrm{s}}$ and $R_{\mathrm{s}}$, respectively, to prevent overlap of the axis labels. Additionally, the LCO 2023, NOT 2023, and NOT 2024 datasets are denoted by the suffixes `L23', `N23', and `N24', respectively.}
    \label{fig:lccorner}
\end{figure*}


\bsp	
\label{lastpage}
\end{document}